\documentclass[9pt,shortpaper,twoside,web]{ieeecolor}
\usepackage{generic}
\usepackage{cite}
\usepackage{amsmath,amssymb,amsfonts}
\usepackage{algorithmic}
\usepackage{graphicx}
\usepackage{textcomp}
\usepackage{float}
\usepackage[table,xcdraw]{xcolor}
\usepackage[acronym]{glossaries}
\usepackage{adjustbox}
\usepackage{svg}
\usepackage{subcaption}

\usepackage{makecell}
\usepackage{comment}

\usepackage{nohyperref}

\definecolor{matplotlib0}{HTML}{1f77b4}
\definecolor{matplotlib1}{HTML}{d62728}
\definecolor{matplotlib2}{HTML}{2ca02c}
\definecolor{matplotlib3}{HTML}{ff7f0e}
\definecolor{matplotlib4}{HTML}{9467bd}
\definecolor{matplotlib5}{HTML}{8c564b}
\definecolor{matplotlib6}{HTML}{e377c2}
\definecolor{matplotlib7}{HTML}{7f7f7f}
\definecolor{matplotlib8}{HTML}{bcbd22}
\definecolor{matplotlib9}{HTML}{17becf}

\usepackage{mathtools} 

\usepackage{subcaption}

\usepackage{booktabs}
\usepackage{multirow}
\usepackage{colortbl}
\usepackage{tablefootnote}
\usepackage{threeparttable}

\usepackage{tikz}
\usetikzlibrary{shapes,arrows,positioning,fit}
\usetikzlibrary{matrix}
\usetikzlibrary{calc}
\usetikzlibrary{external}
\usepackage{adjustbox}

\usepackage{pgfplots}
\definecolor{color0}{rgb}{0.12156862745098,0.466666666666667,0.705882352941177} 
\definecolor{color1}{rgb}{1,0.498039215686275,0.0549019607843137}
\definecolor{color2}{rgb}{0.172549019607843,0.627450980392157,0.172549019607843} 
\definecolor{color3}{rgb}{0.83921568627451,0.152941176470588,0.156862745098039} 
\definecolor{color4}{rgb}{0.580392156862745,0.403921568627451,0.741176470588235}
\definecolor{colorblue}{rgb}{0.12156862745098,0.466666666666667,0.705882352941177} 
\definecolor{colorgreen}{rgb}{0.172549019607843,0.627450980392157,0.172549019607843} 
\definecolor{colorred}{rgb}{0.83921568627451,0.152941176470588,0.156862745098039} 
\definecolor{colorblack}{rgb}{0,0,0} 
\definecolor{amber}{rgb}{1.0, 0.75, 0.0}
\definecolor{cyan(process)}{rgb}{0.0, 0.72, 0.92}
\definecolor{colororange}{rgb}{1,0.56,0} 
\definecolor{colorbrown}{rgb}{0.62, 0.42, 0.21} 
\definecolor{colorpurple}{rgb}{0.33, 0.033, 0.51}
\usepgfplotslibrary{fillbetween}
\usepgfplotslibrary{colormaps}
\pgfplotsset{compat=1.16}

\pgfplotscreateplotcyclelist{matplotlib}{
  {matplotlib0},
  {matplotlib1},
  {matplotlib2},
  {matplotlib3},
  {matplotlib4},
  {matplotlib5},
  {matplotlib6},
  {matplotlib7},
  {matplotlib8},
  {matplotlib9}
}

\pgfplotsset{every axis/.append style={
    cycle list name=matplotlib,
}}

\pgfdeclareplotmark{mystar}{
    \node[star,star point ratio=2.25,minimum size=6pt,
          inner sep=0pt,draw=black,solid,fill=red] {};
}

\usepackage{listings}
\usepgfplotslibrary{groupplots}

\definecolor{code_default}{HTML}{000000}
\definecolor{code_keyword}{HTML}{AC4142}
\definecolor{code_identifier}{HTML}{D28445}

\lstdefinelanguage{RISCV}{
  sensitive=false,
  morecomment=[l]{//},
  alsoletter={.},
  morekeywords=[1]{
    lp.setup, mv, lw, p.lw, sw, p.sw, pv.sdotsp.b, pv.shuffle2.b, p.subNR, p.addNR
  },
  morekeywords=[2]{
    zero, ra, sp, gp, tp, t0, t1, t2, t3, t4, t5, t6, s0, s1, a0, a1, a2, a3, a4, a5, a6, a7, a8, a9, a10, a11,
  },
  morestring=[b]",
  morestring=[b]',
}[strings, comments, keywords]

\lstdefinestyle{RISCV_STYLE}{
  language=RISCV,
  numbers=none,
  basicstyle=\scriptsize\ttfamily\color{code_default},
  keywordstyle=[1]\color{matplotlib0},
  keywordstyle=[2]\color{matplotlib1},
  float,
  captionpos=b,
  belowskip=-0.5cm
}

\def\BibTeX{{\rm B\kern-.05em{\sc i\kern-.025em b}\kern-.08em
    T\kern-.1667em\lower.7ex\hbox{E}\kern-.125emX}}
\begin{document}

\title{FEMBA on the Edge: Physiologically-Aware Pre-Training, Quantization, and Deployment of a Bidirectional Mamba EEG Foundation Model on an Ultra-low Power Microcontroller}

\author{Anna Tegon, \IEEEmembership{Graduate Student Member, IEEE}, Nicholas Lehmann, Yawei Li, Andrea Cossettini, \IEEEmembership{Senior Member, IEEE}, Luca Benini, \IEEEmembership{Fellow, IEEE} and Thorir~Mar~Ingolfsson~\IEEEmembership{Member, IEEE}
\thanks{This project was supported by the Swiss National Science Foundation (Project PEDESITE) under grant agreement 193813. This work was also supported in part by the ETH Future Computing Laboratory (EFCL) and by a grant from the Swiss National Supercomputing Centre (CSCS) under project ID lp12 on Alps.}
\thanks{Anna Tegon, Nicolas Lehmann, Yawei Li, Andrea Cossettini, Luca Benini, and Thorir Mar Ingolfsson are with the Integrated Systems Laboratory, ETH Z{\"u}rich, Z{\"u}rich, Switzerland (thoriri@iis.ee.ethz.ch).}
\thanks{Luca Benini is also with the DEI, University of Bologna, Bologna, Italy.}
}
\makeatletter
\def\ps@mynotice{%
  \def\@oddhead{}%
  \def\@evenhead{}%
  \def\@oddfoot{%
    \hfil
    \parbox[t]{\textwidth}{\centering\scriptsize
      \copyright\ This work has been submitted to the IEEE for possible
      publication. Copyright may be transferred without notice, after which
      this version may no longer be accessible.%
    }%
    \hfil
  }%
  \def\@evenfoot{\@oddfoot}%
}
\makeatother
\maketitle
\thispagestyle{mynotice}   
\newacronym{ofa}{OFA}{Once-For-All}
\newacronym{simd}{SIMD}{Single Instruction, Multiple Data}

\newacronym{elu}{ELU}{Exponential Linear Unit}
\newacronym{relu}{ReLU}{Rectified Linear Unit}
\newacronym{rpr}{RPR}{Random Partition Relaxation}
\newacronym{mac}{MAC}{Multiply Accumulate}
\newacronym{dma}{DMA}{Direct Memory Access}
\newacronym{bmi}{BMI}{Brain--Machine Interface}
\newacronym{bci}{BCI}{Brain--Computer Interface}
\newacronym{smr}{SMR}{Sensory Motor Rythms}
\newacronym{eeg}{EEG}{Electroencephalography}
\newacronym{emg}{EMG}{Electromyography}
\newacronym{svm}{SVM}{Support Vector Machine}
\newacronym{svd}{SVD}{Singular Value Decomposition}
\newacronym{evd}{EVD}{Eigendecomposition}
\newacronym{iir}{IIR}{Infinite Impulse Response}
\newacronym{fir}{FIR}{Finite Impulse Response}
\newacronym{fc}{FC}{Fabric Controller}
\newacronym{nn}{NN}{Neural Network}
\newacronym{mrc}{MRC}{Multiscale Riemannian Classifier}
\newacronym{flop}{FLOP}{Floating Point Operation}
\newacronym{sos}{SOS}{Second-Order Section}
\newacronym{ipc}{IPC}{Instructions per Cycle}
\newacronym{tcdm}{TCDM}{Tightly Coupled Data Memory}
\newacronym{fpu}{FPU}{Floating Point Unit}
\newacronym{fma}{FMA}{Fused Multiply Add}
\newacronym{alu}{ALU}{Arithmetic Logic Unit}
\newacronym{dsp}{DSP}{Digital Signal Processing}
\newacronym{gpu}{GPU}{Graphics Processing Unit}
\newacronym{soc}{SoC}{System-on-Chip}
\newacronym{mi}{MI}{Motor-Imagery}
\newacronym{csp}{CSP}{Commmon Spatial Patterns}
\newacronym{fbcsp}{FBCSP}{Filter-Bank \acrlong{csp}}
\newacronym{pulp}{PULP}{parallel ultra-low power}
\newacronym{soa}{SoA}{state-of-the-art}
\newacronym{bn}{BN}{Batch Normalization}
\newacronym{isa}{ISA}{Instruction Set Architecture}
\newacronym{ecg}{ECG}{Electrocardiogram}
\newacronym{mcu}{MCU}{microcontroller}
\newacronym{rnn}{RNN}{recurrent neural network}
\newacronym{cnn}{CNN}{convolutional neural network}
\newacronym{tcn}{TCN}{temporal convolutional network}
\newacronym{emu}{EMU}{epilepsy monitoring unit}
\newacronym{ml}{ML}{Machine Learning}
\newacronym{dl}{DL}{Deep Learning}
\newacronym{ai}{AI}{Artificial Intelligence}
\newacronym{tcp}{TCP}{Temporal Central Parasagittal}
\newacronym{loocv}{LOOCV}{Leave-One-Out Cross-Validation}
\newacronym{wfcv}{WFCV}{Walk-Forward Cross-Validation}
\newacronym{rwcv}{RWCV}{Rolling Window Cross-Validation}
\newacronym{iot}{IoT}{Internet of Things}
\newacronym{auc}{AUC}{Area Under the Receiver Operator Characteristic}
\newacronym{dwt}{DWT}{Discrete Wavelet Transform}
\newacronym{fft}{FFT}{Fast Fourier Transform}
\newacronym{tpot}{TPOT}{Tree-based Pipeline Optimization Tool}

\newacronym{tuar}{TUAR}{Temple University Artifact Corpus}
\newacronym{tuev}{TUEV}{Temple University Event Corpus}
\newacronym{tuab}{TUAB}{Temple University Abnormal Corpus}
\newacronym{tusl}{TUSL}{Temple University Slowing Corpus}
\newacronym{bss}{BSS}{Blind Source Separation}
\newacronym{ica}{ICA}{Independent Component Analysis}
\newacronym{ic}{ICs}{Independent Components}
\newacronym{asr}{ASR}{Artifact Subspace Reconstruction}
\newacronym{pca}{PCA}{Principal Component Analysis}
\newacronym{gap}{GAP}{Global Average Pooling}
\newacronym{fcn}{FCN}{Fully Connected Networks}
\newacronym{mlp}{MLP}{Multi-Layer Perceptron}
\newacronym{nas}{NAS}{Neural Architectural Search}
\newacronym{fph}{FP/h}{False Positives per Hour}
\newacronym{bvp}{BVP}{Blood volume Pulse}
\newacronym{eda}{EDA}{Electrodermal Activity}
\newacronym{acc}{ACC}{Accelerometer}
\newacronym{cae}{CAE}{Convolutional Autoencoder}
\newacronym{sswce}{SSWCE}{Sensitivity-Specificity Weighted Cross-Entropy}
\newacronym{ce}{CE}{Cross-Entropy}
\newacronym{ppg}{PPG}{plethysmography}
\newacronym{asic}{ASIC}{Application-specific integrated circuit}
\newacronym{sota}{SOTA}{state-of-the-art}
\newacronym{auroc}{AUROC}{Area Under the Receiver Operating Characteristic}
\newacronym{aupr}{AUPR}{Area Under the Precision-Recall}
\newacronym{qat}{QAT}{Quantization Aware Training}
\newacronym{ptq}{PTQ}{Post-Training Quantization}
\newacronym{fm}{FM}{Foundation Model}

\begin{abstract}
\textbf{Objective:} To enable continuous, long-term neuro-monitoring on wearable devices by overcoming the computational bottlenecks of Transformer-based Electroencephalography (EEG) foundation models and the quantization challenges inherent to State-Space Models (SSMs).
\textbf{Methods:} We present FEMBA, a bidirectional Mamba architecture pre-trained on over 21,000 hours of EEG. We introduce a novel Physiologically-Aware pre-training objective, consisting of a reconstruction with low-pass filtering, to prioritize neural oscillations over high-frequency artifacts. To address the activation outliers common in SSMs, we employ Quantization-Aware Training (QAT) to compress the model to 2-bit weights. The framework is deployed on a parallel ultra-low-power RISC-V microcontroller (GAP9) using a custom double-buffered memory streaming scheme.
\textbf{Results:} The proposed low-pass pre-training improves downstream  AUROC on TUAB from 0.863 to 0.893 and  AUPR from 0.862 to 0.898 compared to the best contrastive baseline. QAT successfully compresses weights with negligible performance loss, whereas standard post-training quantization degrades accuracy by approximately \textbf{30\%}. The embedded implementation achieves deterministic real-time inference (\textbf{1.70~s} per 5~s window) and reduces the memory footprint by \textbf{74\%} (to $\approx$2~MB), achieving competitive accuracy with up to \textbf{27$\times$} fewer FLOPs than Transformer benchmarks.
\textbf{Conclusion:} FEMBA demonstrates that Mamba-based foundation models can be effectively quantized and deployed on extreme-edge hardware without sacrificing the representation quality required for robust clinical analysis.
\textbf{Significance:} This work establishes the first full-stack framework for deploying large-scale EEG foundation models on ultra-low-power wearables, facilitating continuous, SSM based monitoring for epilepsy and sleep disorders.
\end{abstract}


\section{Introduction}

\Gls{eeg} measures cortical electrical activity using non-invasive electrodes. Since its earliest use in the 1920s, \gls{eeg} has become a fundamental tool for monitoring brain activity \cite{collura1993history}, with clinical applications spanning the diagnosis of sleep disorders, neurodegenerative diseases, and epilepsy \cite{petit2004sleep,noachtar2009role}.

In recent years, \gls{eeg} applications have been moving outside controlled clinical environments. Clinically, there is a need for continuous, long-term monitoring in ambulatory and at-home settings, for example in the context of epilepsy monitoring \cite{beniczky2021automated}. At the same time, the consumer market shows increased interest towards wellness-oriented brain monitoring solutions and brain computer interfaces (BCI), with applications spanning aided meditation, boosting productivity, and gaming \cite{acabchuk2021measuring,liao2012gaming}.

As \gls{eeg} expands beyond controlled clinical environments, there is a pressing need to enable continuous monitoring and on-device \gls{eeg} signal analysis on resource-constrained wearable devices.
Such systems must operate under tight computational and power constraints~\cite{ahmadi2012brain}, while being more susceptible to movement and environmental artifacts affecting signal quality~\cite{seok2021motion,ingolfsson_minimizing_2024}.

In this context, the algorithmic landscape for \gls{eeg} analysis is rapidly evolving. Classical \gls{eeg} analysis initially relied on handcrafted spectral, temporal, or spatial features combined with traditional classifiers such as Support Vector Machines, Linear Discriminant Analysis, and tree-based methods~\cite{reviewalgorithms}.

Overcoming the limitations of manual feature engineering prompted a shift toward end-to-end learning, with convolutional neural networks~\cite{eegtcnet} enabling automatic feature extraction and improved decoding performance. Despite the increased memory and computational demands, there have been multiple demonstrations of model deployments on edge devices for low-power execution on wearables~\cite{zanetti2021real,ingolfsson_brainfusenet_2024,frey2024gapses}. However, existing approaches still struggle to generalize across subjects, recording platforms, and electrode configurations, often requiring task-specific and patient-specific data collections and training.

\Glspl{fm} address the generalization limitations of deep learning by pre-training on large-scale, unlabeled \gls{eeg} corpora and adapting to downstream tasks through transfer learning. Recent models such as BENDR~\cite{BENDR}, LaBraM~\cite{labram}, and CbraMod~\cite{cbramod} demonstrate improved cross-subject generalization and reduced annotation requirements.
However, while FMs offer the robust generalization capabilities required for diverse patient populations, their reliance on Transformer architectures with quadratic complexity $\mathcal{O}(N^2)$ renders them computationally prohibitive for wearable devices. This disconnect prevents the deployment of state-of-the-art AI on the very edge devices required for long-term, continuous neuro-monitoring.

State-space models (SSMs), such as Mamba~\cite{Mamba}, offer a promising avenue to avoid the fundamental quadratic complexity bottleneck, by reformulating sequence modeling as a latent dynamical system with linear scaling $\mathcal{O}(N)$ in sequence length. Yet, two critical barriers remain for their adoption in biomedical edge computing. First, standard self-supervised pre-training objectives (e.g., masked reconstruction) often force models to reconstruct high-frequency artifacts (e.g., EMG noise), wasting model capacity on non-physiologically meaningful characteristics of the raw signals. Second, Mamba architectures are notoriously difficult to quantize due to activation outliers, so naive integer post-training quantization can lead to severe performance collapse on low-precision \glspl{mcu}.

Building on our previous FEMBA architecture~\cite{tegon2025fembaefficientscalableeeg}, which introduced a bidirectional Mamba-based \gls{eeg} foundation encoder, we present an end-to-end framework that bridges the gap between large-scale \glspl{fm} and ultra–low-power wearable hardware. Our specific contributions are in addition to open-sourcing our code and models for reproducibility\footnote{https://github.com/pulp-bio/BioFoundation}:

\begin{itemize}
 \item \textbf{Physiologically-Aware Pre-training}: We introduce a self-supervised objective, \emph{Reconstruction with Low-pass Filtering}, that acts as a denoising autoencoder by forcing the encoder to reconstruct a low-pass–filtered target instead of raw \gls{eeg}. On \gls{tuab}, this objective improves \gls{auroc} from $0.863$ to $0.893$ and \gls{aupr} from $0.862$ to $0.898$ compared to the best contrastive baseline, with significant accuracy improvement ($78.6\%\rightarrow81.9\%$). On \gls{tuar} and \gls{tusl}, performance differences between the pre-training strategies are within overlapping confidence intervals. The low-pass variant remains competitive and does not compromise performance on these smaller datasets, while consistently avoiding the degradation observed with contrastive masking.

 \item \textbf{Robust Mamba Quantization}: We systematically study post-training quantization (PTQ) and show that activation outliers in Mamba-style SSMs make naive W8A8 PTQ collapse performance (\gls{auroc} $0.89\rightarrow0.77$, accuracy $81\%\rightarrow55\%$), and that W2A8 PTQ fails completely. By switching to Quantization-Aware Training (QAT), we recover near-floating-point performance for W8A8 and W4A8 (within $\approx0.01$ \gls{auroc} of FP32) and even for W2A8\footnote{while activation 4-bit quantization remains unstable in our experiments}. This yields a 2-bit–weight, 8-bit–activation FEMBA-Tiny variant that is both accurate and deployable on tightly constrained \glspl{mcu} where standard PTQ fails.

 \item \textbf{Full-Stack Edge Deployment}: We demonstrate, to the best of our knowledge, the first deployment of a SSM-based \gls{eeg} foundation model on a parallel ultra–low-power RISC-V \gls{mcu}. Using optimized kernels and a double-buffered multi-level memory streaming scheme, we achieve deterministic inference of a 5\,s window in \textbf{1.70\,s} at 370\,MHz, at an energy cost of \textbf{75\,mJ per inference}, at an average power envelope of 44.1 mW, while compressing the model footprint from 7.8\,MB (INT8) to $\sim$2\,MB (2-bit weights). This shows that foundation-scale \gls{eeg} encoders can be made compatible with the memory and latency budgets of continuous \gls{mcu}-based, wearable monitoring devices.
\end{itemize}
These three components—physiologically-aware pre-training, robust quantization, and full-stack deployment—are tightly coupled. The low-pass reconstruction objective stabilizes the encoder’s frequency content, which in turn simplifies the quantization landscape. The resulting quantized model is architected to match the memory hierarchy and processing architecture of \glspl{mcu}.

Compared to our preliminary FEMBA paper~\cite{tegon2025fembaefficientscalableeeg}, which introduced the original bidirectional Mamba encoder and demonstrated the feasibility of pre-training and fine-tuning on \gls{tuab}, \gls{tuar}, and \gls{tusl}, this work makes three key extensions. First, we systematically compare four self-supervised objectives and propose a physiologically-aware low-pass reconstruction target that significantly improves downstream performance on \gls{tuab}. Second, we present, to our knowledge, the first comprehensive quantization study of Mamba for \gls{eeg}, showing that quantization-aware training enables a W2A8 configuration that remains competitive with full-precision baselines. Third, we develop a full-stack deployment framework targeting a low-power \gls{mcu}, including custom kernels, hierarchical streaming, and a detailed cycle-accurate analysis, demonstrating that FEMBA can meet the memory and latency of \gls{mcu}'s highly constrained compute and storage.

\section{Related Work}
\subsection{Supervised Deep Learning for EEG}
Early applications of supervised deep learning to \gls{eeg} primarily relied on convolutional neural networks (CNNs). An important early contribution was the DeepConvNet and ShallowConvNet~\cite{shallowconvnet} architectures, which showed that end-to-end CNNs operating directly on raw \gls{eeg} could match the performance of traditional feature-engineering pipelines. Building on these ideas, EEGNet~\cite{lawhern2018eegnet} introduced a more compact and efficient CNN design based on depthwise-separable convolutions, better capturing the spatiotemporal structure of \gls{eeg} while remaining lightweight and portable. Numerous variants and derived models emerged, including MBEEGNet~\cite{MBEEGNet}, TIDNet~\cite{tidnet} and related architectures, which extend its capabilities through temporal modules, multi-scale processing, and attention mechanisms.
In parallel to CNN-based models, recurrent architectures were explored to better capture the temporal dynamics of \gls{eeg} signals. Early examples include LSTM-based classifiers~\cite{brainwave}, which demonstrated that sequence models applied to sliding windows of raw \gls{eeg} can effectively learn temporal dependencies, and RNN frameworks combined with sliding-window CSP features~\cite{RNN}, which highlighted the potential of recurrent networks for modeling temporal structure in motor-imagery decoding.

\subsection{Foundation Models for EEG Analysis}
Recent foundation models for \gls{eeg} increasingly rely on self-supervised learning (SSL) to exploit large-scale unlabeled recordings. Early work such as BENDR~\cite{BENDR} adapted masked prediction frameworks by combining convolutional encoders with contrastive objectives to reconstruct masked \gls{eeg} representations. Later approaches extended this paradigm using Transformer-based architectures: BrainBERT~\cite{brainbert} performs masked prediction on channel-independent spectrograms for iEEG, while models such as LaBraM~\cite{labram} apply vector-quantized masking and discrete latent spaces to learn robust codebooks. Recent developments, as CBraMod~\cite{cbramod}, reconstruct masked raw signal patches directly, enabling end-to-end learning of temporal and spatial \gls{eeg} structure.

However, a common limitation across these approaches is their agnostic treatment of frequency content. By attempting to reconstruct the full spectral bandwidth, which includes high-frequency noise and muscle artifacts~\cite{ingolfsson_minimizing_2024}, these models may allocate capacity to modeling non-physiological interference rather than cortical dynamics. This suggests an opportunity for physiologically-aware pre-training objectives that prioritize neural oscillations over broadband reconstruction.

Despite the progress of Transformer-based foundation models, their quadratic time and memory complexity with respect to sequence length ($O(n^{2})$) limits their practicality in many real-world \gls{eeg} scenarios, especially in wearable or edge-computing settings where compute and memory resources are constrained~\cite{ingolfsson_brainfusenet_2024}. Applications such as continuous epilepsy monitoring further impose real-time requirements and strict false-alarm tolerances~\cite{ingolfsson_minimizing_2024}.

State-space model (SSM) architectures offer a compelling alternative, as their linear complexity ($\mathcal{O}(N)$) enables efficient processing of long sequences, and recent designs, such as Mamba~\cite{Mamba}, demonstrate strong sequence-modeling performance. While Mamba-based \gls{eeg} models have begun to emerge—such as EEGMamba~\cite{eegmamba} and EEGM2~\cite{hong2025eegm2}, which employ Mixture-of-Experts and U-Net architectures, respectively—these works focus primarily on algorithmic performance on high-end GPUs, leaving the challenges of edge deployment and quantization largely unexplored.

\subsection{Efficient Edge AI and Quantization Challenges}
\gls{eeg} processing is commonly performed offline or on high-performance hardware, and existing surveys mainly focus on acquisition and usability rather than embedded computation. A few recent works started to explore edge-based \gls{eeg} processing, showing that lightweight CNNs can run on \glspl{mcu}, though with strict constraints on memory and latency~\cite{ecg-tcn}. Due to these limitations, several studies have investigated model compression—including pruning, quantization, and compact architectures—to enable deployment on low-power devices~\cite{ingolfsson_brainfusenet_2024}.

However, deploying Foundation Models (specifically SSMs) presents unique challenges compared to standard CNNs. While CNNs are often robust to low-bit quantization (e.g., 8-bit or 4-bit), Mamba architectures are notoriously difficult to quantize. Recent studies in computer vision~\cite{xu2025mambaquant} highlight that Mamba's selective scan mechanism generates activation outliers that destroy performance under standard Post-Training Quantization (PTQ). Consequently, bridging the gap between Mamba's theoretical efficiency and actual hardware implementation requires dedicated Quantization-Aware Training (QAT) strategies that have not yet been applied to the \gls{eeg} domain.

\subsection{Limitations of Prior Works}
In this context, prior works demonstrated the growing interest in self-supervised \gls{eeg} representation learning, yet they reveal clear limitations. As shown in Table~\ref{tab:comparison}, existing foundation models rely on Transformer architectures with quadratic complexity, translating to computational costs up to 27$\times$ higher than linear alternatives. Simultaneously, many \gls{eeg} applications, from Brain-Computer Interfaces to continuous ambulatory monitoring, require online inference on low-power wearable devices. However, the development of foundation models and edge-AI solutions has largely progressed in isolation: as shown in Table~\ref{tab:comparison}, \textbf{none of these models report edge deployment}, leaving the gap between algorithmic performance and embedded feasibility unaddressed. While state-space models such as Mamba~\cite{Mamba} offer linear complexity, their deployment on ultra-low-power devices remains unexplored in the biomedical literature.

We address these gaps by deploying a bidirectional Mamba foundation model for biosignal analysis running on an ultra-low-power edge \gls{mcu}.
 
\begin{table}[h!]
\centering
\caption{Comparison of EEG foundation models}
\label{tab:comparison}
\resizebox{\columnwidth}{!} {
\setlength{\tabcolsep}{2pt} 
\begin{tabular}{@{}lccccccc@{}}
\toprule
\textbf{Model} & \textbf{Architecture} & \textbf{Size} & \textbf{Datasets} & \textbf{Complexity} & \textbf{Deployment} \\
\hline
EEGFormer & Transformer & 2.3M & TUAR / TUSL & $\mathcal{O}(C N^{2})$ & No \\
LaBraM & Transformer & 5.9M & TUAB & $\mathcal{O}(C^{2} N^{2})$ & No \\
LUNA & Transformer & 7M & TUAB / TUAR / TUSL & $\mathcal{O}(C N^{2}) + \mathcal{O}(C N)$ & No \\
FEMBA & Mamba-based & 7.8M & TUAB / TUAR / TUSL & $\mathcal{O}(C N)$ & Yes \\
\hline
\end{tabular}
}
\vspace{0.1cm}\\
{\scriptsize 
$C$  number of EEG channels \\
$N$ number of temporal patches
}
\end{table}
\section{Methods}

\subsection{Datasets}
\label{sec:datasets}
We leveraged the Temple University EEG Corpus (TUEG)~\cite{obeid2016temple} for pretraining, as it is one of the largest publicly available clinical \gls{eeg} repositories. The corpus contains over $21{,}600$ hours of recordings from more than $14{,}000$ patients.
The TUEG dataset includes several labeled subsets designed for specific diagnostic tasks. The \textbf{\gls{tuab}} subset provides recordings labeled as normal or abnormal (binary classification) for $2{,}329$ subjects. The \textbf{\gls{tuar}} comprises data from $213$ subjects and, following prior work ~\cite{tegon2025fembaefficientscalableeeg}, we treat it as a multiclass (single-label) classification task with five classes corresponding to five artifact types. The \textbf{\gls{tusl}} includes recordings from $38$ subjects for the detection and classification of slowing events, seizures, complex backgrounds, and normal \gls{eeg} activity~\cite{obeid2016temple} (multiclass classification).

\begin{table}[htbp]
\caption{Dataset Statistics}
\label{tab:dataset_stats}
\centering
\begin{tabular}{lcccc}
\toprule
\textbf{Property} & \textbf{TUEG} & \textbf{TUAB} & \textbf{TUAR} & \textbf{TUSL} \\
\midrule
Number of Subjects & $14{,}987$ & $2{,}329$ & $213$ & $38$ \\
Number of Channels & 22 & 22 & 22 & 22 \\
Sampling Rate & 256~Hz & 256~Hz & 256~Hz & 256~Hz \\
Hours of Recordings & $21{,}787{.}32$ & $1{,}139{.}31$ & $83{.}74$ & $27{.}54$ \\
Training Samples & $13{,}236{,}000$ & $591{,}357$ & $49{,}241$ & $16{,}088$ \\
Validation Samples & $489{,}600$ & $154{,}938$ & $5{,}870$ & $1{,}203$ \\
Test Samples & $489{,}600$ & $74{,}010$ & $5{,}179$ & $2{,}540$ \\
\bottomrule
\end{tabular}
\end{table}

\subsection{Preprocessing}
\label{subesc:preprocessing}
We applied a standard pre-processing pipeline to the raw \gls{eeg} recordings. All signals were band-pass filtered between 1~Hz and 75~Hz, and a 60~Hz notch filter was used to remove power line interference. Signals were then resampled to 256~Hz for consistency across recordings. After resampling, each raw signal was segmented into non-overlapping 5 s windows for training and evaluation.

An initial analysis of the TUEG dataset showed that most signals (about 96\%) were within the range of --20.16$\mu\mathrm{V}$ to 19.96$\mu\mathrm{V}$, while the remaining recordings contained values with much higher magnitudes. To preserve dataset integrity and ensure comparability with prior work using the full TUEG dataset, we retained all recordings, including those with extreme values. Given $x$, raw \gls{eeg} signal, $x_{\text{norm}}$ corresponds to its normalized version, which is then provided as input to the model.
To reduce the influence of these artifacts during training, we applied quartile-based normalization~\cite{bedeeuzzaman2012automatic}, scaling each channel by its interquartile range (IQR):
\begin{equation*}
  x_{\text{norm}} = \frac{x - q_{\text{lower}}}{(q_{\text{upper}} - q_{\text{lower}}) + 1 \times 10^{-8}}.
\end{equation*}

where $q_{\text{lower}}$ and $q_{\text{upper}}$ denote the 25th and 75th percentiles (i.e., the lower and upper quartiles) of each channel's amplitude distribution, respectively. The small constant $1 \times 10^{-8}$ is added for numerical stability.

\subsection{Model Architecture}
We build on the architecture introduced in our previous work~\cite{tegon2025fembaefficientscalableeeg}. With the goal of enabling model deployment on low-power \glspl{mcu}, we adopt the \emph{FEMBA-Tiny} configuration for all the following experiments. The Tiny variant consists of a 2D convolutional tokenizer that projects the raw \gls{eeg} input into an embedding space of dimension $d_{model}=385$. The encoder is composed of two Bidirectional Mamba (Bi-Mamba) blocks, which process the sequence in both forward and backward directions to capture complex temporal dependencies. A residual connection is included within each Bi-Mamba block to support gradient flow during training.

A key architectural novelty is the introduction of an additional lightweight Transformer layer in the decoder. Specifically, we integrate this layer to enhance the model’s contextual reasoning capabilities. This hybrid design leverages the structured state-space modeling of Mamba with the contextual reasoning ability of self-attention~\cite{lieber2024jamba}. The decoder is used only during pretraining and is discarded in downstream classification, preserving the linear computational complexity of the model during fine-tuning.

In the classification stage, a simple linear classifier is employed, consisting of a single fully connected layer. This minimalistic task-specific output layer highlights the key role of the pretrained encoder.

\subsection{Pre-training}
\label{sec:pretraining}

Recent work on foundational \gls{eeg} models has explored a range of pretraining strategies, showing promise for both masked-reconstruction approaches~\cite{cbramod} and contrastive learning methods~\cite{BENDR}. In this study, we aim to investigate whether one pretraining paradigm consistently yields stronger performance. To this end, we evaluate four pretraining techniques: two based on contrastive learning and two based on masked reconstruction. 
 
We pretrain FEMBA-Tiny on the TUEG dataset (see Sect.~\ref{sec:datasets}). All subjects and recordings contained in the downstream datasets (\gls{tuab}, \gls{tuar}, and \gls{tusl}) were excluded from the pretraining to ensure a fair assessment of the model’s generalization capabilities.

\subsubsection{Masked Reconstruction Approaches}
\label{subsec:reconstructiontask}

In these setups, the input signal is divided into $80$ non-overlapping patches of size $16$. A random subset of these patches is replaced with a fixed mask token using a masking ratio between $0.5$ and $0.6$, consistent with prior work~\cite{labram}. The model is trained to reconstruct the original signal $\hat{x} = f(x_{\text{m}})$ by minimizing the Smooth L1 loss~\cite{girshick2015fastrcnn}:
\begin{equation}
  \text{SmoothL1}(\hat{x}, x) =
  \begin{cases}
    0.5 \, (x - \hat{x})^2 / \beta, & \text{if } |x - \hat{x}| < \beta, \\
    |x - \hat{x}| - 0.5 \beta, & \text{otherwise.}
  \end{cases}
  \label{eq:smoothl1}
\end{equation}
We compute the loss over all patches, weighting unmasked patches by $0.1$ to maintain consistent representations. We evaluate two specific reconstruction targets:

\paragraph{\textbf{Low-pass Filtering}}
While high-frequency components (e.g., HFOs) contain relevant biomarkers, they are frequently contaminated by muscle artifacts (\gls{emg}) in scalp \gls{eeg}~\cite{ingolfsson_minimizing_2024}. For robust ambulatory monitoring, we prioritize low-frequency morphology (0.5--40~Hz)~\cite{arpaia2025systematic}. We apply a 2\textsuperscript{nd}-order biquad low-pass filter (40~Hz cutoff) to the \emph{target} signal only. This introduces an implicit denoising objective, encouraging the model to recover meaningful neural activity while ignoring the high-frequency noise.

\paragraph{\textbf{Clustered Random Patches}}
To increase reconstruction task difficulty, we employ a clustered masking strategy. Instead of independent random masking, we group masked regions into contiguous segments, maintaining the $0.5$ to $0.6$ ratio. This prevents the model from relying on local interpolation and forces it to learn longer-range relations, temporal structure, and cross-channel dependencies.

\subsubsection{Contrastive Learning Approaches}
\label{subsec:contrastive}
We also evaluate contrastive learning, which aims to learn discriminative representations by pulling together positive pairs (augmented views of the same signal) and pushing away negative pairs~\cite{yang2022unsupervised}. We maximize the similarity between views using the standard InfoNCE loss~\cite{oord2018representation}:

\begin{equation}
\mathcal{L}_{\text{InfoNCE}} = - \log 
\frac{\exp(\text{sim}(z_i, z_i^{+}) / \tau)}
{\sum_{j=1}^{N} \exp(\text{sim}(z_i, z_j) / \tau)},
\end{equation}
where $z_i$ and $z_i^{+}$ are embeddings of two views of the same signal, and $\tau$ is the temperature parameter. We explore two view-generation strategies:

\paragraph{\textbf{Frequency-domain Augmentations}}
Following Rommel \textit{et al.}~\cite{rommel2022data}, we make views using three complementary transformations which simulate inter-subject variability and sensor noise: (1) \textit{FT Surrogate}~\cite{schwabedal2018addressing}, which randomizes phase while preserving the magnitude spectrum; (2) \textit{Frequency Shift}~\cite{rommel2021cadda}, which shifts spectral components via the Hilbert transform; and (3) additive Gaussian noise.

\paragraph{\textbf{Masking-based Augmentations}}
Inspired by self-supervised audio learning~\cite{yonay2025myna}, we generate positive pairs by applying two non-overlapping binary masks to the same input signal. Unlike reconstruction, which focuses on waveform details, this objective forces the model to identify latent neural patterns that are semantically consistent across different temporal views of the recording.

 \subsection{Fine-tuning Methodology}
We evaluated \textbf{FEMBA-Tiny} on \gls{eeg}-based classification tasks, specifically three downstream tasks: \textit{epilepsy detection} (\gls{tuab}), \textit{slowing event detection} (\gls{tusl}), and \textit{artifact detection} (\gls{tuar}).

We maintained the same downstream tasks as in our prior work~\cite{tegon2025fembaefficientscalableeeg}, as they provide a broad spectrum of classification scenarios, including both binary and multi-class settings. Moreover, these tasks span different application domains, from clinical diagnostics to artifact identification, requiring the model to adapt to diverse signal characteristics.
For \gls{tuab}, we preserved the standard train-test partition provided with the dataset. For \gls{tusl} and \gls{tuar}, which lack predefined subject-level splits, we followed the evaluation protocol established by recent state-of-the-art methods, including EEGFormer~\cite{eegformer}, applying a randomized 80\%/10\%/10\% division at the sample level for training, validation, and testing to ensure fair and direct comparison with existing benchmark results. For \gls{tuar} specifically, we formulated the problem as a 5-class single-label classification task, focusing on five distinct artifact categories, consistent with prior work.

For the classifier architecture, we removed the decoder and replaced it with a lightweight linear classification head. Thanks to the improvements introduced during pretraining, we were able to eliminate the additional Mamba block used in the previous setup. This change reduced the number of parameters in the classification head from approximately 0.7 million to just a few thousand, significantly simplifying the fine-tuning process.

For the \gls{tuab} dataset, which is relatively balanced, we adopted the standard cross-entropy loss, as it consistently yields stable and robust training performance. Conversely, for the imbalanced \gls{tuar} and \gls{tusl} datasets, we employed the Focal Loss~\cite{lin2017focal} to mitigate class imbalance. The loss function is defined as:

\begin{equation}
\mathrm{FL}(p_t) = -\alpha_t (1 - p_t)^{\gamma} \log(p_t)
\end{equation}

where $p_t$ denotes the predicted probability of the correct class, $\alpha_t$ is a class-balancing factor, and $\gamma$ is a focusing parameter. Focal Loss reduces the impact of well-classified samples and emphasizes harder, minority-class examples, while rebalancing each class contribution by frequency. This approach proved particularly effective for the \gls{tusl} dataset, which exhibits significant class imbalance.

We performed fine-tuning using the AdamW optimizer with a batch size of 256 over 50 epochs. The learning rate was set to $5\text{e-}4$ with a layer-wise learning rate decay of 0.7. Additionally, we applied a weight decay of 0.05 and a gradient clipping threshold of 1.0, while dropout was set to 0.0.

\subsection{Quantization}
\label{quantization_methods}
To quantize FEMBA-Tiny, we utilized \textit{Brevitas}, a PyTorch-based library for neural network quantization that supports both \gls{ptq} and \gls{qat}~\cite{brevitas}. These represent the two general approaches to quantization, depending on when quantization is applied. \gls{ptq} is performed after training, while \gls{qat} is conducted during training~\cite{rokh2023comprehensive}.

\gls{ptq} is the simplest approach for quantizing a pre-trained model. It typically relies on basic assumptions and statistics to perform the quantization. This process can be improved through \textit{calibration}, which in Brevitas is implemented using the \texttt{calibration\_mode} and \texttt{bias\_correction\_mode} functions. These allow collecting activation statistics in floating point with quantization temporarily disabled, and then re-enabling quantization with properly initialized scales, followed by optional bias correction.

On the other hand, \gls{qat} is the most complex yet effective quantization method, as it simulates quantization during training, allowing the model to adapt to quantization effects and reduce accuracy degradation.

In our evaluation of FEMBA-Tiny, we quantized both weights and activations down to 2-bit precision. We use \textbf{per-channel uniform quantization} for the weights. For each output
channel \(c\), a separate scale factor \(s_w^{(c)}\) is learned. The quantization
process is given by:

\[
q_w^{(c)} = \operatorname{round}\!\left(\frac{w^{(c)}}{s_w^{(c)}}\right),
\qquad
\hat{w}^{(c)} = s_w^{(c)} \cdot q_w^{(c)},
\]

where \( w^{(c)} \) denotes the original weights in channel \(c\), 
\( s_w^{(c)} \) is the per-channel floating-point scale, 
\( q_w^{(c)} \) is the quantized integer representation, 
and \( \hat{w}^{(c)} \) is the reconstructed (dequantized) weight.

For activations, instead of using a floating-point scale, we opted for a more hardware-friendly fixed-point quantization by constraining the scale to be a power-of-two value:

\[
s_a = 2^{-n},
\quad
q_a = \text{round}\left(\frac{a}{s_a}\right),
\quad
\hat{a} = s_a \cdot q_a,
\]

where \( a \) is the original activation value, \( s_a \) is the power-of-two scale, \( q_a \) is the quantized activation, and \( \hat{a} \) is the reconstructed activation.

This design choice enables efficient implementation on hardware accelerators by replacing multiplication with bit-shifting operations.
 
\subsection{Embedded deployment}
\label{sec:deployment}
We deploy the quantized FEMBA-Tiny model on the GAP9 RISC-V \gls{mcu}~\cite{greenwaves_gap_sdk} to demonstrate for the first time the feasibility of running Mamba-based foundation models on extreme edge devices. 

\subsubsection{Hardware and Memory Hierarchy}
The GAP9 features a cluster of 9 RISC-V cores (8 workers, 1 orchestrator) with a hierarchical memory architecture: 128~kB L1 scratchpad, 1.5~MB shared L2, and off-chip L3 HyperRAM~\cite{daghero2025lightweight}. We utilize the XpulpV2 ISA extensions, including a 4-way INT8 SIMD dot-product and hardware loops, to accelerate compute-intensive operations.

To handle weights exceeding on-chip capacity, we implemented a \textbf{hierarchical double-buffered streaming strategy}. Large weight matrices are partitioned into $\approx$80~kB chunks and streamed L3$\to$L2 via DMA. Simultaneously, the cluster orchestrator manages L2$\to$L1 tiling, prefetching weights into L1 double-buffers while worker cores compute on the current tile. This hides memory latency and enables deployment of models limited only by external storage size.

\subsubsection{Hybrid Quantization and Kernels}
We employ a \textbf{hybrid quantization strategy} tailored to the numerical requirements of SSMs. We developed a custom toolchain to export parameters directly from Brevitas to optimized C kernels.

\paragraph{\textbf{Linear Projections (INT8)}}
The input/output projections and gate generations dominate the parameter count. We quantize these to INT8 and parallelize execution across the 8 worker cores. The kernels use an output stationary dataflow with $4\times4$ loop unrolling (over outputs and timesteps) to maximize register reuse. The innermost loop utilizes SIMD instructions to perform four MACs per cycle, accumulating into an INT32 to prevent overflow.

\paragraph{\textbf{Selective SSM Scan (Q15)}}
The recurrent scan $h_t = \bar{A} \cdot h_{t-1} + \bar{B} \cdot x_t$ is sensitive to quantization noise. We implement this operation using Q15 fixed-point arithmetic (15-bit fractional precision). Unlike the linear layers, the scan contains sequential temporal dependencies, preventing time-dimension parallelization. Instead, we adopt a \textbf{channel-parallel strategy}: the $d_{inner} = 1540$ channels are distributed across cores, with each core processing its subset sequentially over time. To also avoid expensive runtime floating-point exponentials, the discretizations parameters ($\bar{A}, \bar{B}$) and SiLU activations are precomputed using Lookup Tables (LUTs).

\subsubsection{Deployment Toolchain}
We developed a custom code generation pipeline to deploy FEMBA-Tiny on GAP9. The pipeline extracts quantized weights and activation scales directly from the Brevitas-trained PyTorch model~\cite{brevitas}, bypassing ONNX to maintain precise control over quantization parameters. A template-based C code generator produces layer-specific kernels with tiling configurations optimized for GAP9's memory hierarchy. The Mamba-specific operators (selective scan, gating, bidirectional combination) are implemented as custom kernels integrated into the GAP SDK build system. This approach enables bit-exact reproducibility between the Python reference implementation and the embedded deployment~\cite{garofalo2020pulp}.
\section{Results}

\definecolor{colNormal}{RGB}{147,167,179}
\definecolor{colEyeMovement}{RGB}{50,114,216}
\definecolor{colMuscleMovement}{RGB}{244,82,179}
\definecolor{colElectrodeArtifact}{RGB}{216,116,38}
\definecolor{colChewing}{RGB}{45,133,53}
\definecolor{colShivering}{RGB}{148,103,189}

\begin{figure}[htbp]
\centering
\includegraphics[width=0.48\linewidth]{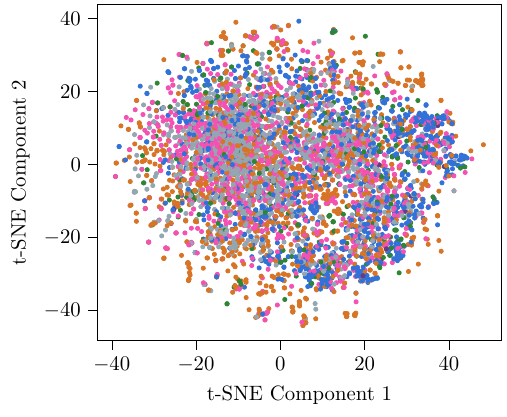}
\hfill
\includegraphics[width=0.48\linewidth]{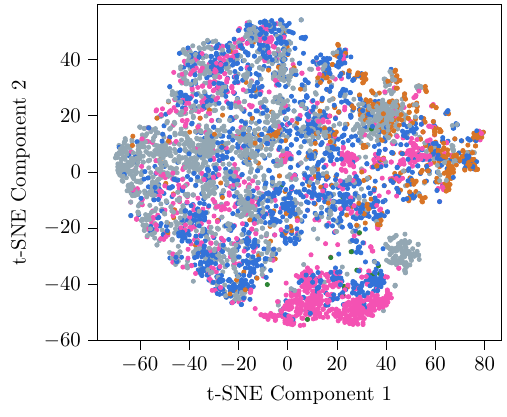}

\vspace{0.5em}
\centering
\footnotesize
\textcolor{colNormal}{$\bullet$}~Normal\quad
\textcolor{colChewing}{$\bullet$}~Chewing\quad
\textcolor{colElectrodeArtifact}{$\bullet$}~Electrode Artifact\\[0.2em]
\textcolor{colEyeMovement}{$\bullet$}~Eye Movement\quad
\textcolor{colMuscleMovement}{$\bullet$}~Muscle Movement\quad
\textcolor{colShivering}{$\bullet$}~Shivering

\caption{t-SNE visualization of embedding spaces on downstream tasks. \textbf{Left}: Embeddings generated using the earlier training pipeline show lower class separability. \textbf{Right}: Embeddings obtained with the updated pipeline yield tighter, more distinct clusters for artifact classes (e.g., muscle movement), illustrating an improvement in the learned representation quality.}
\label{fig:comparison_pretraining}
\end{figure}
\subsection{Pre-training Performance- Comparisons}
\label{pretaining performance comparison}
To identify the optimal pre-training objective, we compared the four proposed strategies across the TUAB, \gls{tuar}, and \gls{tusl} datasets, as summarized in Table~\ref{tab:pretraining_comparison}.

For the smaller \gls{tuar} and \gls{tusl} datasets, the differences between pre-training strategies are modest and fall within overlapping confidence intervals, precluding claims of a superior method.

In contrast, the larger \gls{tuab} dataset reveals distinct performance hierarchies. The \textbf{\textit{Reconstruction--Random with LowPass}} strategy emerges as the superior approach, outperforming all alternatives across every metric. Most notably, it achieves a $>$3.5\% improvement in AUROC compared to the \textit{Contrastive--Frequency} baseline and maintains a $\sim$2\% lead in  \gls{aupr} against all other methods. In terms of accuracy, it remains competitive with the \textit{Clustered Random} variant while surpassing the contrastive approaches by margins of 1--3\%.

\begin{table*}[htbp]
\centering
\caption{Comparison of Fine-tuning Results Across TUAR, TUSL, and TUAB Datasets for Different Pre-training Strategies}
\label{tab:pretraining_comparison}
\resizebox{\textwidth}{!}{
\begin{tabular}{l cc c cc c ccc}
\toprule
\multirow{2}{*}{\textbf{SSL Task}} & \multicolumn{2}{c}{\textbf{TUAR}} & & \multicolumn{2}{c}{\textbf{TUSL}} & & \multicolumn{3}{c}{\textbf{TUAB}} \\
\cmidrule{2-3} \cmidrule{5-6} \cmidrule{8-10}
 & \textbf{AUROC} & \textbf{AUPRC} & & \textbf{AUROC} & \textbf{AUPRC} & & \textbf{Accuracy (\%)} & \textbf{AUROC} & \textbf{AUPRC} \\
\midrule
Rec. -- Random Masking & $0.912\pm 0.002$ & $0.532 \pm 0.011$ & & $0.699 \pm 0.028$ & $0.281 \pm 0.019$ & & $80.38 \pm 0.08$ & $0.8762 \pm 0.0012$ & $0.8773 \pm 0.0011$ \\

Rec. -- Clustered Random & $0.917 \pm 0.004$ & $0.547 \pm 0.027$ & & $0.712 \pm 0.023$ & $0.281 \pm 0.019$ & & $81.38 \pm 0.09$ & $0.8663 \pm 0.001$ & $0.8654 \pm 0.0011$ \\
Rec. -- Random w/ Lowpass & $0.916 \pm 0.008$ & $0.533 \pm 0.024$ & & $0.750 \pm 0.035$ & $0.294 \pm 0.009$ & & $\mathbf{81.95 \pm 0.09}$ & $\mathbf{0.8930 \pm 0.0007}$ & $\mathbf{0.8976 \pm 0.0018}$ \\
Contrastive -- Frequency & $0.921 \pm 0.007$ & $0.534 \pm 0.023$ & & $0.751 \pm 0.020$ & $0.291 \pm 0.014$ & & $78.63 \pm 0.09$ & $0.8633 \pm 0.0007$ & $0.8615 \pm 0.0015$ \\
Contrastive -- Masking & $0.915 \pm 0.012$ & $0.533 \pm 0.014$ & & $0.695 \pm 0.020$ & $0.272 \pm 0.004$ & & $80.69 \pm 0.07$ & $0.8732 \pm 0.0006$ & $0.8691 \pm 0.0007$ \\
\bottomrule
\end{tabular}
}
\end{table*}

Having established \textit{Reconstruction–Random with LowPass} as the most effective pretraining strategy for FEMBA, we also compared the resulting encoder with the previous version of the model \cite{tegon2025fembaefficientscalableeeg}. Since the decoder is discarded during fine-tuning, the downstream performance depends entirely on the quality of the learned representations. As illustrated qualitatively in Fig.~\ref{fig:comparison_pretraining}, the new pipeline produces embeddings that form more coherent class-specific clusters (e.g., distinguishing muscle artifacts from eye movements), indicating a substantial improvement in representation quality.

\subsection{Fine-tuning Performance}

Given the considerations discussed in Section~\ref{pretaining performance comparison}, we adopt the Reconstruction–Random with LowPass configuration as the new pretraining strategy for all downstream tasks.  
In Tables~\ref{tab:results_TUAR_TUSL} and ~\ref{tab:TUAB_sota}, we report results for all downstream evaluations and compare them with the smallest available architecture of the original FEMBA model for each respective dataset, as well as to recent state-of-the-art (SoA) \gls{eeg} foundation models.  
Since the new FEMBA architecture is fine-tuned using the smallest possible classifier (a linear head), whereas the original FEMBA required a heavy Mamba classifier, we ensure fairness by comparing the new FEMBA model against the original pretrained FEMBA encoder fine-tuned with both a Mamba classifier and a linear classifier.

\subsubsection{TUAB dataset}
Compared to the smallest model previously used (FEMBA-Base~\cite{tegon2025fembaefficientscalableeeg}), the updated FEMBA demonstrates improved performance when evaluated against both variants of the original model, i.e., the version fine-tuned with the Mamba classifier and the version fine-tuned with a linear classifier.

With respect to the \textbf{Mamba-classifier} baseline, we observe slightly higher accuracy (around 1\%) and  \gls{aupr} (around 1\%), while obtaining similar \gls{auroc} values. When comparing against the \textbf{Linear-classifier} version of the original model, gains are more substantial: approximately 2\% in accuracy, 4\% in  \gls{auroc}, and up to 5\% in  \gls{aupr}, despite having roughly $\mathbf{6\times}$ fewer parameters.

Overall, the new FEMBA ranks among the strongest models on the benchmark. It is analogous to LaBraM-Base~\cite{labram}, performing marginally better in accuracy (around a 0.5\% improvement) while showing minimally worse  \gls{auroc} and  \gls{aupr} (around 0.5\%). Compared to much larger models such as LaBraM-Huge and CBraMod~\cite{cbramod}, the accuracy drop remains modest (only 0.5--0.6\%), despite their massively larger parameter counts, highlighting the efficiency of the proposed 7.8M-parameter architecture.

Notably, this improved performance is achieved with only 1.3G FLOPs, up to $\mathbf{6\times}$ fewer than FEMBA Base and $\mathbf{27\times}$ fewer than LaBraM-Base.

\subsubsection{TUAR dataset}

 Among the previous FEMBA variants, the smallest directly comparable model is FEMBA-Tiny. When comparing the updated FEMBA to the original Tiny model fine-tuned with the \textbf{Mamba-classifier}, the two models exhibit equivalent  \gls{auroc} performance, as their confidence intervals nearly overlap, while the updated version achieves a modest improvement in  \gls{aupr} (about 2\% on average).
Against the Tiny model fine-tuned with a \textbf{Linear-classifier}, the updated FEMBA shows clearer gains:  \gls{auroc} improves by roughly 2--3\%, and  \gls{aupr} by nearly 6\%, despite operating at essentially the same parameter scale.
In comparison to state-of-the-art models, the updated FEMBA achieves competitive performance. Compared to the much larger LUNA-Huge~\cite{doner2025luna} (SoA  \gls{auroc}), it achieves slightly higher average  \gls{aupr} (on the order of 0.5\%), while exhibiting  \gls{auroc} confidence intervals that closely overlap, thus reaching comparable results with approximately $\mathbf{40\times}$ fewer parameters. When compared to the larger FEMBA-Base (SoA AUC-PR), the updated model reaches similar  \gls{auroc} values while showing lower  \gls{aupr} (around 2--3\%). Overall, these results demonstrate that the updated FEMBA achieves a favorable trade-off between model size and performance across both metrics.
\subsubsection{TUSL dataset}
On the \gls{tusl} dataset, as shown in Table~\ref{tab:results_TUAR_TUSL}, the updated FEMBA demonstrates consistent improvements over previous models of comparable size. Relative to the original Tiny model (the smallest version of the previous FEMBA) equipped with the \textbf{Mamba-classifier},  \gls{auroc} and  \gls{aupr} increase by approximately 4\% and 1.5\%, respectively. 
A comparison with the \textbf{Linear-classifier} variant reveals even larger gains in  \gls{auroc}, around 5\%, while the improvement in  \gls{aupr} remains similar to the previous case.
When benchmarked against leading models, the updated FEMBA maintains a competitive standing. Although LUNA-Huge (SOTA  \gls{auroc}) surpasses it by roughly 5--6\% in  \gls{auroc}, their  \gls{aupr} scores are nearly indistinguishable, with overlapping confidence intervals, despite FEMBA using approximately $\mathbf{40\times}$ fewer parameters. 
Compared to EEGFormer-Base (SOTA  \gls{aupr}), the updated model attains a higher  \gls{auroc} ($\approx$ 3--4\%), but lags in  \gls{aupr} by approximately 10\%.

\begin{table}[H]
\centering
\caption{Comparison of model performance on TUAR and TUSL datasets.}
\resizebox{\columnwidth}{!} {
\setlength{\tabcolsep}{2pt} 
\begin{tabular}{@{}lccccccc@{}}
\toprule
\textbf{Model} & \textbf{Size} & \multicolumn{2}{c}{\textbf{TUAR}} & \multicolumn{2}{c}{\textbf{TUSL}} \\
\cmidrule(lr){3-4} \cmidrule(lr){5-6}
 & & AUROC $\uparrow$ & AUC-PR $\uparrow$ & AUROC $\uparrow$ & AUC-PR $\uparrow$  \\
\midrule
\multicolumn{6}{l}{\textit{Supervised Models}} \\
EEGNet~\cite{lawhern2018eegnet} & - & $0.752 \pm 0.006$ & $0.433 \pm 0.025$ & $0.635 \pm 0.015$ & $0.351 \pm 0.006$ \\
EEG-GNN~\cite{eeggnn} & - & $0.837 \pm 0.022$ & $0.488 \pm 0.015$ & $0.721 \pm 0.009$ & $0.381 \pm 0.004$ \\
GraphS4mer~\cite{graphsmer} & - & $0.833 \pm 0.006$ & $0.461 \pm 0.024$ & $0.632 \pm 0.017$ & $0.359 \pm 0.001$ \\
\midrule
\multicolumn{6}{l}{\textit{Self-supervised Models}} \\
BrainBERT~\cite{brainbert} & 43.2M & $0.753 \pm 0.012$ & $0.350 \pm 0.014$ & $0.588 \pm 0.013$ & $0.352 \pm 0.003$ \\
EEGFormer-Base~\cite{eegformer} & 2.3M & $0.847 \pm 0.014$ & $0.483 \pm 0.026$ & $0.713 \pm 0.010$ & $\mathbf{0.393 \pm 0.003}$ \\
EEGFormer-Large~\cite{eegformer} & 3.2M & $0.852 \pm 0.004$ & $0.483 \pm 0.014$ & $0.679 \pm 0.013$ & $0.389 \pm 0.003$ \\
FEMBA-Base~\cite{tegon2025fembaefficientscalableeeg} & 47.7M & $0.900 \pm 0.010$ & $0.559 \pm 0.002$ & $0.731 \pm 0.012$ & $0.289 \pm 0.009$ \\
FEMBA-Large~\cite{tegon2025fembaefficientscalableeeg} & 77.8M & $0.915 \pm 0.003$ & $0.521 \pm 0.001$ & $0.714 \pm 0.007$ & $0.282 \pm 0.010$ \\
LUNA-Base ~\cite{doner2025luna}  & 7M & $0.902 \pm 0.011$ & $0.495 \pm 0.010$ & $\mathbf{0.767 \pm 0.023}$ & $0.301 \pm 0.003$ \\
LUNA-Huge & 311.4M & $0.921 \pm 0.011$ & $0.528 \pm 0.012$ & $0.802 \pm 0.005$ & $0.289 \pm 0.008$ \\
\midrule

FEMBA old - Mamba classifier\cite{tegon2025fembaefficientscalableeeg} & 8.5M$^{*}$ & $\mathbf{0.918 \pm 0.003}$ & $0.518 \pm 0.002$ & $0.708 \pm 0.005$ & $0.277 \pm 0.007$\\
FEMBA old  - Linear classifier& 7.8M$^{*}$ &$0.893 \pm 0.021$ & $0.475 \pm 0.025$ &$0.688 \pm 0.030$ &$0.272 \pm 0.007$ \\
\textbf{FEMBA new - Linear classifier}& 7.8M$^{*}$ &  $0.916 \pm 0.008$ & $\mathbf{0.533 \pm 0.024}$ &$0.750 \pm 0.035$ & $0.294 \pm 0.009$ \\
\bottomrule
\end{tabular}
}
\vspace{0.01cm}\\
{\scriptsize 
$^{*}$ Model size including the classification head \\
$^{\dagger}$ \textbf{Bold} indicates state-of-the-art models under 10M parameters.
}
\label{tab:results_TUAR_TUSL}
\end{table}

\begin{table}[h!]
\centering
\caption{Performance comparison on \gls{tuab} abnormal EEG detection.}
\label{tab:TUAB_sota}
\resizebox{\columnwidth}{!} {
\setlength{\tabcolsep}{2pt} 
\begin{tabular}{@{}lccccccc@{}}
\toprule
\textbf{Model} & \textbf{Size} & \textbf{Bal. Acc. (\%)} $\uparrow$ & \textbf{AUC-PR} $\uparrow$ & \textbf{AUROC} $\uparrow$ \\
\midrule
\multicolumn{5}{l}{\textit{Supervised Models}} \\
SPaRCNet \cite{sparcnet} & 0.8M & 78.96 $\pm$ 0.18 & 0.8414 $\pm$ 0.0018 & 0.8676 $\pm$ 0.0012 \\
ContraWR \cite{ContraWR} & 1.6M & 77.46 $\pm$ 0.41 & 0.8421 $\pm$ 0.0140 & 0.8456 $\pm$ 0.0074 \\
CNN-Transformer \cite{CNNtrasnformer} & 3.2M & 77.77 $\pm$ 0.22 & 0.8433 $\pm$ 0.0039 & 0.8461 $\pm$ 0.0013 \\
FFCL \cite{FFCL} & 2.4M & 78.48 $\pm$ 0.38 & 0.8448 $\pm$ 0.0065 & 0.8569 $\pm$ 0.0051 \\
ST-Transformer \cite{STtransformer} & 3.2M & 79.66 $\pm$ 0.23 & 0.8521 $\pm$ 0.0026 & 0.8707 $\pm$ 0.0019 \\
\midrule
\multicolumn{5}{l}{\textit{Self-supervised Models}} \\
BENDR \cite{BENDR} & 0.39M & 76.96 $\pm$ 3.98 & - & 0.8397 $\pm$ 0.0344 \\
BrainBERT \cite{brainbert} & 43.2M & - & 0.8460 $\pm$ 0.0030 & 0.8530 $\pm$ 0.0020 \\
EEGFormer-Base \cite{eegformer} & 2.3M & - & 0.8670 $\pm$ 0.0020 & 0.8670 $\pm$ 0.0030 \\
BIOT \cite{biot} & 3.2M & 79.59 $\pm$ 0.57 & 0.8692 $\pm$ 0.0023 & 0.8815 $\pm$ 0.0043 \\
EEG2Rep \cite{eeg2rep} & - & 80.52 $\pm$ 2.22 & - & 0.8843 $\pm$ 0.0309 \\

CEREbRO \cite{cerebro} & 85.15M & 81.67 $\pm$ 0.23 & 0.9049 $\pm$ 0.0026 & 0.8916 $\pm$ 0.0038 \\
LaBraM-Base \cite{labram} & 5.9M & 81.40 $\pm$ 0.19 & $\mathbf{0.8965 \pm 0.0016}$ & $\mathbf{0.9022 \pm0.0009}$ \\
LaBraM-Huge \cite{labram} & 369.8M & 82.58 $\pm$ 0.11 & 0.9204 $\pm$ 0.0011 & 0.9162 $\pm$ 0.0016 \\
CBraMod \cite{cbramod} & 69.3M & 82.49 $\pm$ 0.25 &0.9221 $\pm$ 0.0015 & 0.9156 $\pm$ 0.0017 \\
\midrule
 FEMBA old - Mamba classifier\cite{tegon2025fembaefficientscalableeeg}  & 48.3M$^{*}$  &81.05 $\pm$ 0.14 &0.8894 $\pm$0.0050  & 0.8829 $\pm$ 0.0021  \\
FEMBA old - Linear classifier & 47.6M$^{*}$  &79.75 $\pm$ 0.15 &0.8511 $\pm$ 0.0011 &0.8456 $\pm$ 0.0011  \\
\textbf{FEMBA new - Linear classifier} & 7.8M$^{*}$  & $\mathbf{81.95 \pm 0.09}$  &0.8930 $\pm$ 0.0007 &0.8976 $\pm$ 0.0018 \\
\bottomrule
\end{tabular}
}
\vspace{0.01cm}\\
{\scriptsize 
$^{*}$ Model size including the classification head \\
$^{\dagger}$ \textbf{Bold} indicates state-of-the-art models under 10M parameters.
}
\end{table}

\subsection{Quantization Analysis}
We focus our quantization analysis on \gls{tuab}, being the most extensively explored dataset in the literature~\cite{lee2025comprehensive}. \gls{tuab} also offers a controlled evaluation setting thanks to its balanced binary classification task, which allows us to isolate the effects of quantization more reliably. Table~\ref{tab:quantization} summarizes the quantization results.

We first evaluated the behavior of the model quantizing only the weights to 8-bit, 4-bit, and 2-bit. The 8-bit as well as 4-bit quantization resulted in negligible performance degradation (0.1\%  \gls{auroc} for 8-bit and 0.1\% for both  \gls{auroc} and  \gls{aupr} for 4-bit). However, reducing to 2-bit led to a substantial degradation in accuracy,  \gls{auroc} and  \gls{aupr}.
Quantizing activations, even at 8-bit precision, yielded an accuracy drop of approximately 30\%, resulting in performance comparable to random guessing (accuracy $\approx$ 0.5). We extensively explored PTQ calibration pipelines in an attempt to mitigate this degradation. Following the procedure detailed in Section~\ref{quantization_methods}, we performed activation calibration in floating point with quantization temporarily disabled, using a dedicated calibration split constructed to be class-balanced and representative of both seizure and non-seizure windows. We experimented with calibration sets ranging from approximately 5\% to 20\% of the \gls{tuab} training windows, confirming that neither increasing the calibration set size nor enforcing class balance produced any measurable improvement. After collecting activation statistics, we re-enabled quantization with the calibrated scales and applied bias correction, yet performance remained essentially unchanged.

This persistent failure of PTQ aligns with recent analyses of Mamba and other state-space models, which report that these architectures naturally generate large activation outliers, particularly in gate and output projections and their associated matrix multiplications. The parallel scan operation further amplifies these effects, resulting in heavy-tailed activation distributions that are inherently difficult to capture with low-precision quantization~\cite{xu2025mambaquant}.
Given the relatively small model size, we were able to apply \gls{qat}. After a few epochs, \gls{qat} recovered the original  \gls{auroc},  \gls{aupr}, and accuracy for weight-only 8-bit and for weight 4-bit with 8-bit activations. Even in the 2-bit weight and 8-bit activation setting, despite the significant drop with PTQ in  \gls{auroc} and  \gls{aupr}, \gls{qat} effectively restored performance, reaching results comparable to those obtained for weight 8-bit or weight 4-bit. However, when quantizing activations to 4-bit, \gls{qat} was not sufficient to recover performance.

\begin{table}[htpb]
\centering
\caption{Quantization Results: Comparison of PTQ and \gls{qat} Performance}
\label{tab:quantization}
\scriptsize
\setlength{\tabcolsep}{3pt}
\renewcommand{\arraystretch}{0.95}
\begin{tabular}{lcccc}
\toprule
\textbf{Configuration} & \textbf{Method} & \textbf{AUROC} & \textbf{AUPR} & \textbf{Accuracy} \\
\midrule
\textbf{FP32 (Baseline)} & -- & 0.89 & 0.89 & 81.84 \\
\midrule
\textbf{W8A8} & PTQ & 0.77 & 0.71 & 55.91 \\
              & QAT & 0.88 & 0.88 & 81.02 \\
\textbf{W4A8} & PTQ & 0.71 & 0.67 & 55.79 \\
              & QAT & 0.88 & 0.88 & 80.88 \\
\textbf{W2A8} & PTQ & 0.56 & 0.49 & 54.12 \\
              & QAT & 0.88 & 0.88 & 80.61 \\
\midrule
\textbf{W4A4} & PTQ & 0.68 & 0.63 & 54.95 \\
              & QAT & 0.69 & 0.68 & 65.40 \\
\textbf{W2A4} & PTQ & 0.54 & 0.54 & 48.38 \\
              & QAT & 0.55 & 0.55 & 49.20 \\
\midrule
\textbf{Only Weight 8} & PTQ & 0.88 & 0.89 & 81.61 \\
\textbf{Only Weight 4} & PTQ & 0.88 & 0.88 & 80.86 \\
\textbf{Only Weight 2} & PTQ & 0.54 & 0.48 & 54.71 \\
\bottomrule
\end{tabular}
\vspace{-0.3cm}
\end{table}
In the remainder of this work we therefore focus on a QAT-trained \emph{FEMBA-Tiny} model with a W2A8 configuration for the encoder weights and activations, combined with a small number of higher-precision accumulators where required for numerical stability. On GAP9, this scheme primarily reduces the model's memory footprint, from 7.8\,MB for a uniform INT8 implementation to approximately 2\,MB, while leaving the runtime dominated by the sequential SSM scan. We use the corresponding W8A8 model as our on-device INT8 baseline. In other words, quantization is crucial to make FEMBA-Tiny fit within the L3/L2 memory budgets of the device and to enable efficient streaming, whereas further reductions in latency will require architectural changes to the SSM itself rather than more aggressive bitwidth scaling. In the next subsection, we deploy both the W8A8 and W2A8 FEMBA-Tiny models on GAP9 and characterize their end-to-end latency, energy, and cycle breakdown.

\subsection{Model deployment on a low-power microcontroller}
We now evaluate the QAT-trained \emph{FEMBA-Tiny} models on a GAP9-based ultra-low-power \gls{mcu} platform. In particular, we compare the uniform INT8 (W8A8) baseline against the 2-bit weight variant (W2A8) selected in the quantization analysis and quantify the impact of 2-bit weights on latency, energy, and memory footprint. As discussed above, both models share the same network architecture and differ only in the numerical representation of the encoder weights and the associated packing/unpacking logic; all activations remain 8-bit on device.

\subsubsection{Experimental Setup}
We evaluated FEMBA-Tiny on a GAP9 development board~\cite{greenwaves_gap_sdk,ingolfsson_minimizing_2024} with the compute cluster operating at 370~MHz. Performance measurements use the GAP9 hardware performance counters to obtain cycle-accurate timing and instruction counts. Each inference processes a 5~s input window with $22$ channels and $1{,}280$ timesteps (256~Hz sampling), represented as a tensor of shape $(22, 1280)$, with embedding dimension $d_{model}=385$. All reported metrics are averaged over 10 runs with identical inputs to ensure measurement stability. MACs/cycle values refer to dense-equivalent INT8 multiply-accumulate operations.

Table~\ref{tab:gap9_perf} summarizes the end-to-end execution metrics for the standard INT8 implementation and the 2-bit weight quantization variant.

\subsubsection{Latency and Efficiency}
The INT8 deployment achieves an inference latency of 1.70 seconds for a 5-second input window (629.4 million cycles), which in turn yields a real-time slack factor of approximately 3x. At the measured average power of 44.1~mW, this corresponds to an energy cost of 75~mJ per 5~s inference window (1.70~s of active compute). At the \gls{tusl} sampling rate this corresponds to roughly $3\times$ faster-than-real-time processing, leaving ample slack for on-device pre- and post-processing. Our double-buffered streaming architecture successfully hides nearly all memory transfer latency, achieving 99.4--100\% overlap between computation and L3$\rightarrow$L2 DMA transfers in the Mamba blocks.

To put these numbers into a more practical perspective, for a typical 300~mAh, 3.7~V wearable battery (around 4.0~kJ of stored energy), this energy cost would allow on the order of $5.3 \times 10^{4}$ such 5~s inference windows, corresponding to roughly 3~days of continuous operation for FEMBA inference alone, ignoring the additional overhead of sensing, storage, and wireless communication. While a full device-level power budget is beyond the scope of this work, these estimates suggest that \textit{foundation-scale \gls{eeg} encoders can be integrated into wearable neuro-monitoring systems without violating typical battery and form-factor constraints.}

As Table~\ref{tab:gap9_layer} shows, the two bidirectional Mamba blocks account for 98.3\% of total cycles, while other layers, patch embedding, positional encoding, global pooling, and classification contribute negligibly to latency. This concentration of runtime in the Mamba blocks confirms that they are the primary target for further optimization.

\subsubsection{Sub-operation Analysis and Efficiency}
To localize bottlenecks within the Mamba blocks, we instrumented each sub-operation. Table~\ref{tab:gap9_subop} reports the breakdown for the first Mamba block (\texttt{mamba\_blocks.0}).

\paragraph{Memory Bandwidth Saturation}
Our hierarchical memory management strategy proves highly effective for the compute-intensive layers. The Input and Output Projections achieve high computational density, reaching 2.65 and 3.91\,MACs/cycle, respectively. This performance confirms that our multi-level double-buffered tiling successfully eliminates memory bandwidth bottlenecks: despite the Input Projection requiring the streaming of 1.13\,MB of weights from off-chip L3 memory, the DMA/Compute overlap remains $>$99\%, ensuring that the cores are never stalled waiting for data.

\paragraph{Shift from MACs/Cycle to IPC}
In contrast, the Selective SSM Scan dominates execution time (64.6\% of cycles) while contributing only a small fraction of the total MACs. This discrepancy highlights a critical distinction for deploying State Space Models: unlike CNNs or Transformers, where performance scales with arithmetic throughput (MACs/cycle), \textit{efficiency in Mamba-based models is defined by \gls{ipc}}

To validate this, we analyzed hardware performance counters. The SSM scan achieves a high \textbf{IPC of 1.36}, indicating that the GAP9 dual-issue pipeline is fully utilized. However, the recurrence inherently requires approximately \textbf{4.3 supporting instructions} (pointer arithmetic, LUT-based parameter discretization, and state management) per MAC operation. Therefore, low MAC utilization in the SSM scan is not a sign of inefficiency, but a characteristic of the algorithm. We conclude that future hardware-software optimization for SSMs must pivot away from maximizing MAC density and instead toward handling complex, scalar instruction streams at high \gls{ipc}.

\subsubsection{Impact of 2-bit Quantization}
We compared the INT8 baseline against 2-bit weight quantization (Table~\ref{tab:gap9_perf}). While 2-bit quantization reduces the model storage by 74\% ($\approx$2\,MB vs.\ 7.8\,MB), end-to-end latency remains nearly unchanged (1.69\,s vs.\ 1.70\,s).

The 2-bit weights are packed four values per byte using a ternary encoding ($\{-1, 0, +1\} \mapsto \{0, 1, 2\}$), yielding 16 weights per 32-bit word. During inference, weights are unpacked on the fly within the dot-product loop: each iteration extracts four 2-bit values via shift-and-mask operations, maps them back to $\{-1, 0, +1\}$ as INT8, and feeds them to the same SIMD dot-product instructions used by the INT8 kernel. This adds roughly eight simple ALU operations per eight weights, which is negligible compared to the memory-access and accumulation costs, explaining why latency is nearly unchanged despite the 4$\times$ reduction in weight memory traffic~\cite{garofalo2020pulp,rutishauser2024xtern}. This behavior is consistent with prior work on PULP-class MCUs, where sub-byte quantization mainly improves model size and energy efficiency, while 8-bit SIMD remains throughput-optimal in the absence of dedicated low-bit instructions~\cite{garofalo2020pulp,rutishauser2024xtern}.

These results confirm that FEMBA-Tiny is compute-bound rather than memory-bound on GAP9. The 2-bit variant is therefore primarily beneficial for reducing external-memory energy and storage footprint on resource-constrained wearables.

\subsubsection{Accuracy}
The embedded implementation achieves bit-exact consistency with the Python simulation. We validated the output of every layer, confirming a bit-to-bit exact implementation for both the INT8 and 2-bit configurations.

\begin{table}[t]
\caption{FEMBA-Tiny Performance on GAP9 @ 370 MHz}
\label{tab:gap9_perf}
\centering
\begin{tabular}{lcc}
\toprule
\textbf{Metric} & \textbf{INT8 (W8A8)} & \textbf{2-bit (W2A8)} \\
\midrule
Total Cycles & 629.4 M & 625.9 M \\
Inference Time & 1.70 s & 1.69 s \\
Compute Utilization & 98.6\% & 98.7\% \\
Idle/Overhead & 0.5\% & 0.5\% \\
DMA/Compute Overlap & 99.4\% & 99.5\% \\
\bottomrule
\end{tabular}
\end{table}

\begin{table}[t]
\caption{Layer-by-Layer Cycle Breakdown}
\label{tab:gap9_layer}
\centering
\begin{tabular}{lrrr}
\toprule
\textbf{Layer} & \textbf{Cycles (M)} & \textbf{\% Total} & \textbf{Overlap} \\
\midrule
patch\_embed & 8.0 & 1.3\% & 81.1\% \\
pos\_embed & 2.3 & 0.4\% & 100.0\% \\
mamba\_blocks.0 & 310.3 & 49.3\% & 100.0\% \\
mamba\_blocks.1 & 308.2 & 49.0\% & 99.4\% \\
global\_pool & 0.5 & 0.1\% & 100.0\% \\
classifier & $<$0.1 & $<$0.01\% & 41.9\% \\
\midrule
\textbf{Total} & \textbf{629.4} & \textbf{100\%} & --- \\
\bottomrule
\end{tabular}
\end{table}

\begin{table}[t]
\caption{Sub-operation Breakdown for \texttt{mamba\_blocks.0}}
\label{tab:gap9_subop}
\resizebox{\columnwidth}{!} {
\centering
\begin{tabular}{lrrrr}
\toprule
\textbf{Operation} & \textbf{Cycles (M)} & \textbf{\%} & \textbf{MACs (M)} & \textbf{MACs/Cyc} \\
\midrule
Input Projection & 71.7 & 23.2\% & 189.7 & 2.65 \\
Sequence Reversal & 1.6 & 0.5\% & --- & --- \\
Local Temporal Conv. & 8.8 & 2.8\% & 1.0 & 0.11 \\
Selective SSM Scan & 199.5 & 64.6\% & 63.1 & 0.32 \\
Output Projection & 24.3 & 7.9\% & 94.9 & 3.91 \\
Sequence Reversal & 1.0 & 0.3\% & --- & --- \\
Bidirectional Fusion & 2.1 & 0.7\% & --- & --- \\
\midrule
\textbf{Total} & \textbf{308.9} & \textbf{100\%} & \textbf{348.7} & \textbf{1.13} \\
\bottomrule
\end{tabular}
}
\end{table}
\section{Discussion}
\subsection{Physiological Awareness in Foundation Models}
Our results demonstrate that the \textit{Reconstruction--Random with LowPass} strategy significantly outperforms standard masking and contrastive approaches. We hypothesize that this objective functions as a domain-specific regularizer. Standard masked modeling forces the network to allocate capacity to reconstructing high-frequency components, which in scalp \gls{eeg} are often dominated by electromyographic (EMG) artifacts and environmental noise \cite{goncharova2003emg}. By filtering the reconstruction target, we explicitly direct the model's attention toward the delta--beta bands ($0.5$--$30$\,Hz), which contain the majority of clinically relevant biomarkers for seizure and artifact detection. This suggests that for biomedical time-series, "faithful" reconstruction of the raw signal is suboptimal; instead, reconstruction targets should be aligned with the physiological bandwidth of interest.

\subsection{The Quantization-Efficiency Trade-off}
The successful compression of FEMBA to 2-bit weights (W2A8) without performance collapse is a key finding. Prior work on Mamba has highlighted the difficulty of \gls{ptq} due to activation outliers in the selective scan\cite{xu2025mambaquant}. Our experiments confirm this: \gls{ptq} failed catastrophically (\gls{auroc} $\approx 0.5$). However, \gls{qat} allowed the model to adapt its weights to the low-precision regime. 
Interestingly, while 2-bit quantization reduced the memory footprint by 74\% (enabling deployment on the 1.5MB L2 memory of GAP9), it did not significantly reduce latency. This confirms that our implementation is \textit{compute-bound}, not memory-bound. The implications are two-fold: (1) aggressive quantization is essential for \textit{storage} on MCUs, but (2) accelerating inference requires dedicated hardware support for the SSM scan operations, rather than just memory bandwidth reduction.

\subsection{Limitations and Future Work}
While FEMBA demonstrates robust performance across several adult \gls{eeg} datasets, important limitations exist. First, our pre-training and evaluation are confined to the TUEG, \gls{tuab}, \gls{tuar}, and \gls{tusl} corpora, which primarily comprise clinical adult recordings from related acquisition pipelines. Generalization to substantially different domains—such as neonatal \gls{eeg} with burst–suppression patterns, high-density research montages, consumer headsets with few dry electrodes, or home-based long-term monitoring—remains unestablished. Extending the pre-training to more diverse data sources and explicitly quantifying out-of-distribution robustness will be crucial.

Second, the current deployment operates on fixed, non-overlapping 5s windows and performs offline classification. Many clinical and consumer applications, however, require streaming or event-triggered processing with strict low-latency constraints. Adapting FEMBA to a fully streaming inference regime—e.g., by exploiting causal variants of the state-space recurrence and reusing internal states across windows—could reduce latency and energy further, but may require retraining and task-specific calibration.

Third, our quantization scheme is validated primarily on the FEMBA-Tiny architecture and \gls{tuab}-based downstream tasks. Although the W2A8 configuration with a small number of higher-precision accumulators proved sufficient in this setting, different hyperparameters, model scales, or target \glspl{mcu} may exhibit different sensitivity to quantization noise. Systematically exploring mixed-precision assignments and automatically tuning quantization parameters for new hardware platforms is, therefore, an open direction.

Fourth, this study focuses on algorithmic and embedded feasibility rather than clinical workflow integration. All evaluations are performed on retrospective datasets; we do not assess prospective performance, user comfort, or the impact of FEMBA-based alerts on clinical decision-making. Future work must investigate how such models affect false-alarm rates, time-to-detection, and interpretability in real-world monitoring scenarios, ideally with clinical partners.

Finally, our hardware analysis highlights a fundamental architectural bottleneck: the selective SSM scan achieves only about 0.32 MACs/cycle on the GAP9 cluster due to its inherently sequential nature and limited instruction-level parallelism, whereas dense linear projections can reach up to 3.9 MACs/cycle. Recent formulations such as Mamba-2, which recast state-space recurrences into structured matrix multiplications, offer a promising path to replace this scan with matmul-friendly kernels that better exploit the GAP9 compute fabric. Given the strong efficiency of our linear layers, integrating such “matmul-friendly” SSM variants is a natural avenue to further accelerate edge inference in future versions of FEMBA.

\section{Conclusion}
This work bridges the gap between large-scale Foundation Models and the resource constraints of wearable biomedical devices. We introduced FEMBA, a bidirectional Mamba architecture that leverages a novel Physiologically-Aware pre-training strategy to prioritize the reconstruction of neural oscillations over high-frequency artifacts. This approach yields superior generalization on diverse downstream tasks (\gls{tuab}, \gls{tuar}, \gls{tusl}) compared to standard masked modeling.

Furthermore, we addressed the critical challenge of deploying State-Space Models on \glspl{mcu}. By utilizing \gls{qat}, we overcame the activation outlier issue inherent to Mamba, successfully compressing the model to 2-bit weights with negligible performance degradation. The resulting deployment on a parallel RISC-V \gls{mcu} (GAP9) achieves deterministic real-time inference (1.70~s per window) with a 74\% reduction in memory footprint.

These results demonstrate that the linear complexity of SSMs, combined with quantization, makes them a viable alternative to Transformer-based models for ambulatory neuro-monitoring. By enabling high-performance artifact and seizure detection directly at the edge, FEMBA paves the way for energy-efficient, long-term wearable health systems that do not rely on continuous cloud connectivity. Future work will focus on architectural approximations to further parallelize the selective scan mechanism, unlocking the full throughput potential of embedded parallel clusters.

\bibliographystyle{IEEEtran}
\bibliography{bib}

@inproceedings{tegon2025fembaefficientscalableeeg,
	title        = {FEMBA: Efficient and Scalable EEG Analysis with a Bidirectional Mamba Foundation Model},
	author       = {Tegon, Anna and Ingolfsson, Thorir Mar and Wang, Xiaying and Benini, Luca and Li, Yawei},
	year         = 2025,
	booktitle    = {2025 47th Annual International Conference of the IEEE Engineering in Medicine and Biology Society (EMBC)},
	volume       = {},
	number       = {},
	pages        = {1--7},
	doi          = {10.1109/EMBC58623.2025.11252697}
}

@inproceedings{girshick2015fastrcnn,
	title        = {Fast {R-CNN} {ICCV}},
	author       = {Girshick, Ross},
	year         = 2015,
	booktitle    = {Proc. of the IEEE Int. Conf. on Computer Vision (ICCV)},
	pages        = {1440--1448}
}

@article{bedeeuzzaman2012automatic,
	title        = {Automatic seizure detection using inter quartile range},
	author       = {Bedeeuzzaman, M and Farooq, Omar and Khan, Yusuf Uzzaman},
	year         = 2012,
	journal      = {Int. Journal of Computer Applications},
	publisher    = {Foundation of Computer Science},
	volume       = 44,
	number       = 11,
	pages        = {1--5}
}

@article{obeid2016temple,
	title        = {The {Temple} {University} {Hospital} {EEG} {Data} {Corpus}},
	author       = {Obeid, Iyad and Picone, Joseph},
	year         = 2016,
	month        = may,
	journal      = {Frontiers in Neuroscience},
	volume       = 10,
	doi          = {10.3389/fnins.2016.00196},
	issn         = {1662-453X},
	language     = {English}
}

@article{lieber2024jamba,
	title        = {Jamba: A hybrid transformer-mamba language model},
	author       = {Lieber, Opher and Lenz, Barak and Bata, Hofit and Cohen, Gal and Osin, Jhonathan and Dalmedigos, Itay and Safahi, Erez and Meirom, Shaked and Belinkov, Yonatan and Shalev-Shwartz, Shai and others},
	year         = 2024,
	journal      = {arXiv preprint arXiv:2403.19887}
}

@inproceedings{labram,
	title        = {Large Brain Model for Learning Generic Representations with Tremendous {EEG} Data in {BCI}},
	author       = {Weibang Jiang and Liming Zhao and Bao-liang Lu},
	year         = 2024,
	booktitle    = {The Twelfth International Conference on Learning Representations}
}

@article{oord2018representation,
	title        = {Representation learning with contrastive predictive coding},
	author       = {Oord, Aaron van den and Li, Yazhe and Vinyals, Oriol},
	year         = 2018,
	journal      = {arXiv preprint arXiv:1807.03748}
}

@inproceedings{yang2022unsupervised,
	title        = {Unsupervised time-series representation learning with iterative bilinear temporal-spectral fusion},
	author       = {Yang, Ling and Hong, Shenda},
	year         = 2022,
	booktitle    = {International conference on machine learning},
	pages        = {25038--25054},
	organization = {PMLR}
}

@article{rommel2022data,
	title        = {Data augmentation for learning predictive models on {EEG}: a systematic comparison},
	author       = {Rommel, C{\'e}dric and Paillard, Joseph and Moreau, Thomas and Gramfort, Alexandre},
	year         = 2022,
	journal      = {Journal of Neural Engineering},
	publisher    = {IOP Publishing},
	volume       = 19,
	number       = 6,
	pages        = {066020}
}

@article{schwabedal2018addressing,
	title        = {Addressing class imbalance in classification problems of noisy signals by using {Fourier} transform surrogates},
	author       = {Schwabedal, Justus TC and Snyder, John C and Cakmak, Ayse and Nemati, Shamim and Clifford, Gari D},
	year         = 2018,
	journal      = {arXiv preprint arXiv:1806.08675}
}

@inproceedings{rommel2021cadda,
	title        = {{CADDA}: Class-wise Automatic Differentiable Data Augmentation for {EEG} Signals},
	author       = {C{\'e}dric Rommel and Thomas Moreau and Joseph Paillard and Alexandre Gramfort},
	year         = 2022,
	booktitle    = {International Conference on Learning Representations}
}

@article{yonay2025myna,
	title        = {Myna: Masking-Based Contrastive Learning of Musical Representations},
	author       = {Yonay, Ori and Hammond, Tracy and Yang, Tianbao},
	year         = 2025,
	journal      = {arXiv preprint arXiv:2502.12511}
}

@inproceedings{lin2017focal,
	title        = {Focal loss for dense object detection},
	author       = {Lin, Tsung-Yi and Goyal, Priya and Girshick, Ross and He, Kaiming and Doll{\'a}r, Piotr},
	year         = 2017,
	booktitle    = {Proceedings of the IEEE international conference on computer vision},
	pages        = {2980--2988}
}

@article{rokh2023comprehensive,
	title        = {A comprehensive survey on model quantization for deep neural networks in image classification},
	author       = {Rokh, Babak and Azarpeyvand, Ali and Khanteymoori, Alireza},
	year         = 2023,
	journal      = {ACM Transactions on Intelligent Systems and Technology},
	publisher    = {ACM New York, NY},
	volume       = 14,
	number       = 6,
	pages        = {1--50}
}

@software{brevitas,
	title        = {Xilinx/brevitas},
	author       = {Franco, Giuseppe and Pappalardo, Alessandro and Fraser, Nicholas J},
	year         = 2025,
	publisher    = {Zenodo},
	doi          = {10.5281/zenodo.3333552},
	url          = {https://doi.org/10.5281/zenodo.3333552}
}

@article{sparcnet,
	title        = {Development of expert-level classification of seizures and rhythmic and periodic patterns during {EEG} interpretation},
	author       = {Jing, Jin and Ge, Wendong and Hong, Shenda and Fernandes, Marta Bento and Lin, Zhen and Yang, Chaoqi and An, Sungtae and Struck, Aaron F and Herlopian, Aline and Karakis, Ioannis and others},
	year         = 2023,
	journal      = {Neurology},
	publisher    = {Lippincott Williams \& Wilkins Hagerstown, MD},
	volume       = 100,
	number       = 17,
	pages        = {e1750--e1762}
}

@article{ContraWR,
	title        = {Self-supervised electroencephalogram representation learning for automatic sleep staging: model development and evaluation study},
	author       = {Yang, Chaoqi and Xiao, Cao and Westover, M Brandon and Sun, Jimeng},
	year         = 2023,
	journal      = {JMIR AI},
	publisher    = {JMIR Publications Inc., Toronto, Canada},
	volume       = 2,
	number       = 1,
	pages        = {e46769}
}

@inproceedings{CNNtrasnformer,
	title        = {Transformer convolutional neural networks for automated artifact detection in scalp {EEG}},
	author       = {Peh, Wei Yan and Yao, Yuanyuan and Dauwels, Justin},
	year         = 2022,
	booktitle    = {2022 44th Annual International Conference of the IEEE Engineering in Medicine \& Biology Society (EMBC)},
	pages        = {3599--3602},
	organization = {IEEE}
}

@article{FFCL,
	title        = {Motor imagery {EEG} classification algorithm based on {CNN-LSTM} feature fusion network},
	author       = {Li, Hongli and Ding, Man and Zhang, Ronghua and Xiu, Chunbo},
	year         = 2022,
	journal      = {Biomedical signal processing and control},
	publisher    = {Elsevier},
	volume       = 72,
	pages        = 103342
}

@article{STtransformer,
	title        = {Transformer-based spatial-temporal feature learning for {EEG} decoding},
	author       = {Song, Yonghao and Jia, Xueyu and Yang, Lie and Xie, Longhan},
	year         = 2021,
	journal      = {arXiv preprint arXiv:2106.11170}
}

@article{BENDR,
	title        = {BENDR: Using transformers and a contrastive self-supervised learning task to learn from massive amounts of {EEG} data},
	author       = {Kostas, Demetres and Aroca-Ouellette, Stephane and Rudzicz, Frank},
	year         = 2021,
	journal      = {Frontiers in Human Neuroscience},
	publisher    = {Frontiers Media SA},
	volume       = 15,
	pages        = 653659
}

@inproceedings{brainbert,
	title        = {Brain{BERT}: Self-supervised representation learning for intracranial recordings},
	author       = {Christopher Wang and Vighnesh Subramaniam and Adam Uri Yaari and Gabriel Kreiman and Boris Katz and Ignacio Cases and Andrei Barbu},
	year         = 2023,
	booktitle    = {The Eleventh International Conference on Learning Representations}
}

@inproceedings{eegformer,
	title        = {{EEGF}ormer: Towards Transferable and Interpretable Large-Scale {EEG} Foundation Model},
	author       = {Yuqi Chen and Kan Ren and Kaitao Song and Yansen Wang and Yifan Wang and Dongsheng Li and Lili Qiu},
	year         = 2024,
	booktitle    = {AAAI 2024 Spring Symposium on Clinical Foundation Models}
}

@article{biot,
	title        = {Biot: Biosignal transformer for cross-data learning in the wild},
	author       = {Yang, Chaoqi and Westover, M and Sun, Jimeng},
	year         = 2023,
	journal      = {Advances in Neural Information Processing Systems},
	volume       = 36,
	pages        = {78240--78260}
}

@inproceedings{eeg2rep,
	title        = {Eeg2rep: enhancing self-supervised {EEG} representation through informative masked inputs},
	author       = {Mohammadi Foumani, Navid and Mackellar, Geoffrey and Ghane, Soheila and Irtza, Saad and Nguyen, Nam and Salehi, Mahsa},
	year         = 2024,
	booktitle    = {Proceedings of the 30th ACM SIGKDD Conference on Knowledge Discovery and Data Mining},
	pages        = {5544--5555}
}

@article{cerebro,
	title        = {{CEReBrO}: Compact encoder for representations of brain oscillations using efficient alternating attention},
	author       = {Dimofte, Alexandru and Bucagu, Glenn Anta and Ingolfsson, Thorir Mar and Wang, Xiaying and Cossettini, Andrea and Benini, Luca and Li, Yawei},
	year         = 2025,
	journal      = {arXiv preprint arXiv:2501.10885}
}

@inproceedings{cbramod,
	title        = {{CB}raMod: A Criss-Cross Brain Foundation Model for {EEG} Decoding},
	author       = {Jiquan Wang and Sha Zhao and Zhiling Luo and Yangxuan Zhou and Haiteng Jiang and Shijian Li and Tao Li and Gang Pan},
	year         = 2025,
	booktitle    = {The Thirteenth International Conference on Learning Representations}
}

@article{lawhern2018eegnet,
	title        = {{EEGNet}: a compact convolutional neural network for {EEG}-based brain--computer interfaces},
	author       = {Lawhern, Vernon J and Solon, Amelia J and Waytowich, Nicholas R and Gordon, Stephen M and Hung, Chou P and Lance, Brent J},
	year         = 2018,
	journal      = {Journal of neural engineering},
	publisher    = {iOP Publishing},
	volume       = 15,
	number       = 5,
	pages        = {056013}
}

@inproceedings{eeggnn,
	title        = {{EEG-GNN}: Graph neural networks for classification of electroencephalogram ({EEG}) signals},
	author       = {Demir, Andac and Koike-Akino, Toshiaki and Wang, Ye and Haruna, Masaki and Erdogmus, Deniz},
	year         = 2021,
	booktitle    = {2021 43rd Annual International Conference of the IEEE Engineering in Medicine \& Biology Society (EMBC)},
	pages        = {1061--1067},
	organization = {IEEE}
}

@inproceedings{graphsmer,
	title        = {Modeling multivariate biosignals with graph neural networks and structured state space models},
	author       = {Tang, Siyi and Dunnmon, Jared A and Liangqiong, Qu and Saab, Khaled K and Baykaner, Tina and Lee-Messer, Christopher and Rubin, Daniel L},
	year         = 2023,
	booktitle    = {Conference on health, inference, and learning},
	pages        = {50--71},
	organization = {PMLR}
}

@inproceedings{doner2025luna,
	title        = {{LUNA}: Efficient and Topology-Agnostic Foundation Model for {EEG} Signal Analysis},
	author       = {Berkay D{\"o}ner and Thorir Mar Ingolfsson and Luca Benini and Yawei Li},
	year         = 2025,
	booktitle    = {The Thirty-ninth Annual Conference on Neural Information Processing Systems}
}

@inproceedings{xu2025mambaquant,
	title        = {MambaQuant: Quantizing the Mamba Family with Variance Aligned Rotation Methods},
	author       = {Zukang Xu and Yuxuan Yue and Xing Hu and Dawei Yang and Zhihang Yuan and Zixu Jiang and Zhixuan Chen and JiangyongYu and XUCHEN and Sifan Zhou},
	year         = 2025,
	booktitle    = {The Thirteenth International Conference on Learning Representations}
}

@article{lee2025comprehensive,
	title        = {A Comprehensive Review of Biosignal Foundation Models},
	author       = {Lee, Na and Barmpas, Konstantinos and Koliousis, Alexandros and Panagakis, Yannis and Adamos, Dimitrios and Laskaris, Nikolaos and Zafeiriou, Stefanos},
	year         = 2025,
	journal      = {Authorea Preprints},
	publisher    = {Authorea}
}

@inproceedings{eegtcnet,
	title        = {{EEG-TCNet}: An Accurate Temporal Convolutional Network for Embedded Motor-Imagery Brain–Machine Interfaces},
	author       = {Ingolfsson, Thorir Mar and Hersche, Michael and Wang, Xiaying and Kobayashi, Nobuaki and Cavigelli, Lukas and Benini, Luca},
	year         = 2020,
	booktitle    = {2020 IEEE International Conference on Systems, Man, and Cybernetics (SMC)},
	volume       = {},
	number       = {},
	pages        = {2958--2965},
	doi          = {10.1109/SMC42975.2020.9283028}
}

@article{MBEEGNet,
	title        = {A multibranch of convolutional neural network models for electroencephalogram-based motor imagery classification},
	author       = {Altuwaijri, Ghadir Ali and Muhammad, Ghulam},
	year         = 2022,
	journal      = {Biosensors},
	publisher    = {MDPI},
	volume       = 12,
	number       = 1,
	pages        = 22
}

@article{tidnet,
	title        = {Thinker invariance: enabling deep neural networks for {BCI} across more people},
	author       = {Kostas, Demetres and Rudzicz, Frank},
	year         = 2020,
	journal      = {Journal of Neural Engineering},
	publisher    = {IOP Publishing},
	volume       = 17,
	number       = 5,
	pages        = {056008}
}

@article{shallowconvnet,
	title        = {Deep learning with convolutional neural networks for EEG decoding and visualization},
	author       = {Schirrmeister, Robin Tibor and Springenberg, Jost Tobias and Fiederer, Lukas Dominique Josef and Glasstetter, Martin and Eggensperger, Katharina and Tangermann, Michael and Hutter, Frank and Burgard, Wolfram and Ball, Tonio},
	year         = 2017,
	journal      = {Human brain mapping},
	publisher    = {Wiley Online Library},
	volume       = 38,
	number       = 11,
	pages        = {5391--5420}
}

@article{brainwave,
	title        = {Brain wave classification using long short-term memory network based OPTICAL predictor},
	author       = {Kumar, Shiu and Sharma, Alok and Tsunoda, Tatsuhiko},
	year         = 2019,
	journal      = {Scientific reports},
	publisher    = {Nature Publishing Group UK London},
	volume       = 9,
	number       = 1,
	pages        = 9153
}

@article{RNN,
	title        = {Exploring spatial-frequency-sequential relationships for motor imagery classification with recurrent neural network},
	author       = {Luo, Tian-jian and Zhou, Chang-le and Chao, Fei},
	year         = 2018,
	journal      = {BMC bioinformatics},
	publisher    = {Springer},
	volume       = 19,
	number       = 1,
	pages        = 344
}

@article{ingolfsson_brainfusenet_2024,
	title        = {{BrainFuseNet}: {Enhancing} {Wearable} {Seizure} {Detection} {Through} {EEG}-{PPG}-{Accelerometer} {Sensor} {Fusion} and {Efficient} {Edge} {Deployment}},
	shorttitle   = {{BrainFuseNet}},
	author       = {Ingolfsson, Thorir Mar and Wang, Xiaying and Chakraborty, Upasana and Benatti, Simone and Bernini, Adriano and Ducouret, Pauline and Ryvlin, Philippe and Beniczky, S\'{a}ndor and Benini, Luca and Cossettini, Andrea},
	year         = 2024,
	month        = aug,
	journal      = {IEEE Transactions on Biomedical Circuits and Systems},
	volume       = 18,
	number       = 4,
	pages        = {720--733},
	doi          = {10.1109/TBCAS.2024.3395534},
	issn         = {1940-9990}
}

@article{ingolfsson_minimizing_2024,
	title        = {Minimizing artifact-induced false-alarms for seizure detection in wearable {EEG} devices with gradient-boosted tree classifiers},
	author       = {Ingolfsson, Thorir Mar and Benatti, Simone and Wang, Xiaying and Bernini, Adriano and Ducouret, Pauline and Ryvlin, Philippe and Beniczky, Sandor and Benini, Luca and Cossettini, Andrea},
	year         = 2024,
	month        = feb,
	journal      = {Sci Rep},
	volume       = 14,
	number       = 1,
	pages        = 2980,
	doi          = {10.1038/s41598-024-52551-0},
	issn         = {2045-2322},
	copyright    = {2024 The Author(s)},
	abstract     = {Electroencephalography (EEG) is widely used to monitor epileptic seizures, and standard clinical practice consists of monitoring patients in dedicated epilepsy monitoring units via video surveillance and cumbersome EEG caps. Such a setting is not compatible with long-term tracking under typical living conditions, thereby motivating the development of unobtrusive wearable solutions. However, wearable EEG devices present the challenges of fewer channels, restricted computational capabilities, and lower signal-to-noise ratio. Moreover, artifacts presenting morphological similarities to seizures act as major noise sources and can be misinterpreted as seizures. This paper presents a combined seizure and artifacts detection framework targeting wearable EEG devices based on Gradient Boosted Trees. The seizure detector achieves nearly zero false alarms with average sensitivity values of \.27{\textbackslash}\%\$\$for 182 seizures from the CHB-MIT dataset and \.26{\textbackslash}\%\$\$for 25 seizures from the private dataset with no preliminary artifact detection or removal. The artifact detector achieves a state-of-the-art accuracy of \.95{\textbackslash}\%\$\$(on the TUH-EEG Artifact Corpus dataset). Integrating artifact and seizure detection significantly reduces false alarms--up to \{\textbackslash}\%\$\$compared to standalone seizure detection. Optimized for a Parallel Ultra-Low Power platform, these algorithms enable extended monitoring with a battery lifespan reaching 300 h. These findings highlight the benefits of integrating artifact detection in wearable epilepsy monitoring devices to limit the number of false positives.}

@article{eegmamba,
	title        = {{EEGMamba}: Bidirectional state space model with mixture of experts for {EEG} multi-task classification},
	author       = {Gui, Yiyu and Chen, MingZhi and Su, Yuqi and Luo, Guibo and Yang, Yuchao},
	year         = 2024,
	journal      = {arXiv preprint arXiv:2407.20254}
}

@inproceedings{Mamba,
	title        = {Mamba: Linear-time sequence modeling with selective state spaces},
	author       = {Gu, Albert and Dao, Tri},
	year         = 2024,
	booktitle    = {First conference on language modeling}
}

@article{hong2025eegm2,
	title        = {Eegm2: An efficient mamba-2-based self-supervised framework for long-sequence {EEG} modeling},
	author       = {Hong, Jiazhen and Mackellar, Geoffrey and Ghane, Soheila},
	year         = 2025,
	journal      = {arXiv preprint arXiv:2502.17873}
}

@misc{greenwaves_gap_sdk,
	title        = {{GAP SDK}: SDK for GreenWaves Technologies' GAP8 IoT Application Processor},
	author       = {{GreenWaves Technologies}},
	year         = 2022,
	urldate      = {2025-11-20},
	note         = {Version 4.12.0},
	howpublished = {\url{https://github.com/GreenWaves-Technologies/gap\_sdk}}
}

@article{reviewalgorithms,
	title        = {A review of classification algorithms for EEG-based brain--computer interfaces},
	author       = {Lotte, Fabien and Congedo, Marco and L{\'e}cuyer, Anatole and Lamarche, Fabrice and Arnaldi, Bruno},
	year         = 2007,
	journal      = {Journal of neural engineering},
	publisher    = {IOP Publishing},
	volume       = 4,
	number       = 2,
	pages        = {R1}
}

@inproceedings{daghero2025lightweight,
	title        = {Lightweight Software Kernels and Hardware Extensions for Efficient Sparse Deep Neural Networks on Microcontrollers},
	author       = {Francesco Daghero and Daniele Jahier Pagliari and Francesco Conti and Luca Benini and Massimo Poncino and Alessio Burrello},
	year         = 2025,
	booktitle    = {Eighth Conference on Machine Learning and Systems}
}

@article{garofalo2020pulp,
	title        = {PULP-NN: Accelerating quantized neural networks on parallel ultra-low-power RISC-V processors},
	author       = {Garofalo, Angelo and Rusci, Manuele and Conti, Francesco and Rossi, Davide and Benini, Luca},
	year         = 2020,
	journal      = {Philosophical Transactions of the Royal Society A},
	publisher    = {The Royal Society Publishing},
	volume       = 378,
	number       = 2164,
	pages        = 20190155
}

@inproceedings{rutishauser2024xtern,
	title        = {xtern: Energy-efficient ternary neural network inference on risc-v-based edge systems},
	author       = {Rutishauser, Georg and Mihali, Joan and Scherer, Moritz and Benini, Luca},
	year         = 2024,
	booktitle    = {2024 IEEE 35th International Conference on Application-specific Systems, Architectures and Processors (ASAP)},
	pages        = {206--213},
	organization = {IEEE}
}

@article{goncharova2003emg,
	title        = {EMG contamination of EEG: spectral and topographical characteristics},
	author       = {Goncharova, Irina I and McFarland, Dennis J and Vaughan, Theresa M and Wolpaw, Jonathan R},
	year         = 2003,
	journal      = {Clinical neurophysiology},
	publisher    = {Elsevier},
	volume       = 114,
	number       = 9,
	pages        = {1580--1593}
}

@article{collura1993history,
	title        = {History and evolution of electroencephalographic instruments and techniques},
	author       = {Collura, Thomas F},
	year         = 1993,
	journal      = {Journal of clinical neurophysiology},
	publisher    = {LWW},
	volume       = 10,
	number       = 4,
	pages        = {476--504}
}

@article{petit2004sleep,
	title        = {Sleep and quantitative EEG in neurodegenerative disorders},
	author       = {Petit, Dominique and Gagnon, Jean-Fran{\c{c}}ois and Fantini, Maria Livia and Ferini-Strambi, Luigi and Montplaisir, Jacques},
	year         = 2004,
	journal      = {Journal of psychosomatic research},
	publisher    = {Elsevier},
	volume       = 56,
	number       = 5,
	pages        = {487--496}
}

@article{noachtar2009role,
	title        = {The role of EEG in epilepsy: a critical review},
	author       = {Noachtar, Soheyl and R{\'e}mi, Jan},
	year         = 2009,
	journal      = {Epilepsy \& Behavior},
	publisher    = {Elsevier},
	volume       = 15,
	number       = 1,
	pages        = {22--33}
}

@article{beniczky2021automated,
	title        = {Automated seizure detection using wearable devices: a clinical practice guideline of the International League Against Epilepsy and the International Federation of Clinical Neurophysiology},
	author       = {Beniczky, S{\'a}ndor and Wiebe, Samuel and Jeppesen, Jesper and Tatum, William O and Brazdil, Milan and Wang, Yuping and Herman, Susan T and Ryvlin, Philippe},
	year         = 2021,
	journal      = {Clinical Neurophysiology},
	publisher    = {Elsevier},
	volume       = 132,
	number       = 5,
	pages        = {1173--1184}
}

@article{acabchuk2021measuring,
	title        = {Measuring meditation progress with a consumer-grade EEG device: Caution from a randomized controlled trial},
	author       = {Acabchuk, Rebecca L and Simon, Mareyna A and Low, Spencer and Brisson, Julie M and Johnson, Blair T},
	year         = 2021,
	journal      = {Mindfulness},
	publisher    = {Springer},
	volume       = 12,
	number       = 1,
	pages        = {68--81}
}

@article{liao2012gaming,
	title        = {Gaming control using a wearable and wireless EEG-based brain-computer interface device with novel dry foam-based sensors},
	author       = {Liao, Lun-De and Chen, Chi-Yu and Wang, I-Jan and Chen, Sheng-Fu and Li, Shih-Yu and Chen, Bo-Wei and Chang, Jyh-Yeong and Lin, Chin-Teng},
	year         = 2012,
	journal      = {Journal of neuroengineering and rehabilitation},
	publisher    = {Springer},
	volume       = 9,
	number       = 1,
	pages        = 5
}

@inproceedings{ahmadi2012brain,
	title        = {Brain-computer interface signal processing algorithms: A computational cost vs. accuracy analysis for wearable computers},
	author       = {Ahmadi, Ali and Dehzangi, Omid and Jafari, Roozbeh},
	year         = 2012,
	booktitle    = {2012 Ninth International Conference on Wearable and Implantable Body Sensor Networks},
	pages        = {40--45},
	organization = {IEEE}
}

@article{seok2021motion,
	title        = {Motion artifact removal techniques for wearable EEG and PPG sensor systems},
	author       = {Seok, Dongyeol and Lee, Sanghyun and Kim, Minjae and Cho, Jaeouk and Kim, Chul},
	year         = 2021,
	journal      = {Frontiers in Electronics},
	publisher    = {Frontiers Media SA},
	volume       = 2,
	pages        = 685513
}

@article{zanetti2021real,
	title        = {Real-time EEG-based cognitive workload monitoring on wearable devices},
	author       = {Zanetti, Renato and Arza, Adriana and Aminifar, Amir and Atienza, David},
	year         = 2021,
	journal      = {IEEE transactions on biomedical engineering},
	publisher    = {IEEE},
	volume       = 69,
	number       = 1,
	pages        = {265--277}
}

@article{frey2024gapses,
	title        = {GAPses: Versatile smart glasses for comfortable and fully-dry acquisition and parallel ultra-low-power processing of EEG and EOG},
	author       = {Frey, Sebastian and Lucchini, Mattia Alberto and Kartsch, Victor and Ingolfsson, Thorir Mar and Bernardi, Andrea Helga and Segessenmann, Michael and Osieleniec, Jakub and Benatti, Simone and Benini, Luca and Cossettini, Andrea},
	year         = 2024,
	journal      = {IEEE Transactions on Biomedical Circuits and Systems},
	publisher    = {IEEE}
}

@article{arpaia2025systematic,
  title={A Systematic Review of Techniques for Artifact Detection and Artifact Category Identification in Electroencephalography from Wearable Devices},
  author={Arpaia, Pasquale and De Luca, Matteo and Di Marino, Lucrezia and Duran, Dunja and Gargiulo, Ludovica and Lanteri, Paola and Moccaldi, Nicola and Nalin, Marco and Picciafuoco, Mauro and Robbio, Rachele and others},
  year={2025},
  journal={Sensors},
  publisher={MDPI},
  volume={25},
  number={18},
  pages={5770},
}

@inproceedings{ecg-tcn,
  author={Ingolfsson, Thorir Mar and Wang, Xiaying and Hersche, Michael and Burrello, Alessio and Cavigelli, Lukas and Benini, Luca},
  booktitle={2021 IEEE 3rd International Conference on Artificial Intelligence Circuits and Systems (AICAS)}, 
  title={ECG-TCN: Wearable Cardiac Arrhythmia Detection with a Temporal Convolutional Network}, 
  year={2021},
  volume={},
  number={},
  pages={1-4},
  doi={10.1109/AICAS51828.2021.9458520}}

\end{document}